\documentclass[journal]{IEEEtran}

\usepackage{amssymb}
\usepackage{amsmath,amsfonts}
\usepackage{algorithmic}
\usepackage{algorithm}
\usepackage{array}
\usepackage{tikz}
\usetikzlibrary{matrix}
\usepackage{textcomp}
\usepackage{stfloats}
\usepackage{url}
\usepackage{svg}
\usepackage{verbatim}
\usepackage{graphicx}
\usepackage{subfigure}
\usepackage{multirow}
\usepackage{multicol}
\usepackage[switch]{lineno}
\usepackage{color}

\newcommand{\red}[1]{\textcolor{red}{#1}}
\usepackage{booktabs}
\usepackage[colorlinks,linkcolor=blue]{hyperref}

\begin{document}
\title{An Interpretable Two-Stage Feature Decomposition Method for Deep Learning-based SAR ATR}

\author{Chenwei Wang, \IEEEmembership{Member,~IEEE,} 
        Renjie Xu, \IEEEmembership{Member,~IEEE,}
        Congwen Wu, \IEEEmembership{Student Member,~IEEE,} 
        Cunyi Yin, \IEEEmembership{Member,~IEEE,}
        Ziyun Liao,
        Deqing Mao, \IEEEmembership{Member,~IEEE,}
        Sitong Zhang,
        Hong Yan, \IEEEmembership{Fellow,~IEEE}

    \thanks{
    C. Wang, R. Xu, C. Yin, Z. Liao, S. Zhang, H. Yan are with the Center for Intelligent Multidimensional Data Analysis, City University of Hong Kong, Hong Kong.
    
    C. Wu, D. Mao are with the Department of Electrical Engineering, University of Electronic Science and Technology of China, Chengdu 611731, China. (\emph{Corresponding author: Deqing Mao.})
}}

\maketitle

\begin{abstract}
Synthetic aperture radar automatic target recognition (SAR ATR) has seen significant performance improvements with deep learning. However, the black-box nature of deep SAR ATR introduces low confidence and high risks in decision-critical SAR applications, hindering practical deployment. 
To address this issue, deep SAR ATR should provide an interpretable reasoning basis $r_b$ and logic $\lambda_w$, forming the reasoning logic $\sum_{i} {{r_b^i} \times {\lambda_w^i}} =pred$ behind the decisions.
Therefore, this paper proposes a physics-based two-stage feature decomposition method for interpretable deep SAR ATR, which transforms uninterpretable deep features into attribute scattering center components (ASCC) with clear physical meanings. First, ASCCs are obtained through a clustering algorithm. 
To extract independent physical components from deep features, we propose a two-stage decomposition method. In the first stage, a feature decoupling and discrimination module separates deep features into approximate ASCCs with global discriminability. In the second stage, a multilayer orthogonal non-negative matrix tri-factorization (MLO-NMTF) further decomposes the ASCCs into independent components with distinct physical meanings. The MLO-NMTF elegantly aligns with the clustering algorithms to obtain ASCCs.
Finally, this method ensures both an interpretable reasoning process and accurate recognition results. Extensive experiments on four benchmark datasets confirm its effectiveness, showcasing the method's interpretability, robust recognition performance, and strong generalization capability.
\end{abstract}


\IEEEpeerreviewmaketitle

\section{Introduction}\label{introduction}

\IEEEPARstart{A}{s} a versatile remote sensing technology, synthetic aperture radar (SAR) is widely used in various applications, providing high-resolution images in any weather conditions \cite{intro1, intro2, wang2023entropy}. SAR automatic target recognition (ATR) has been developed over the past fifty years and can be classified into template-, model-, and deep learning-based methods. 
In the last decade, deep learning-based SAR ATR (DL-SAR ATR) methods have significantly improved performance \cite{sar_sparse_network_cls, wang2023sar}.

\begin{figure}[htb]
\centering
\includegraphics[width=0.45\textwidth]{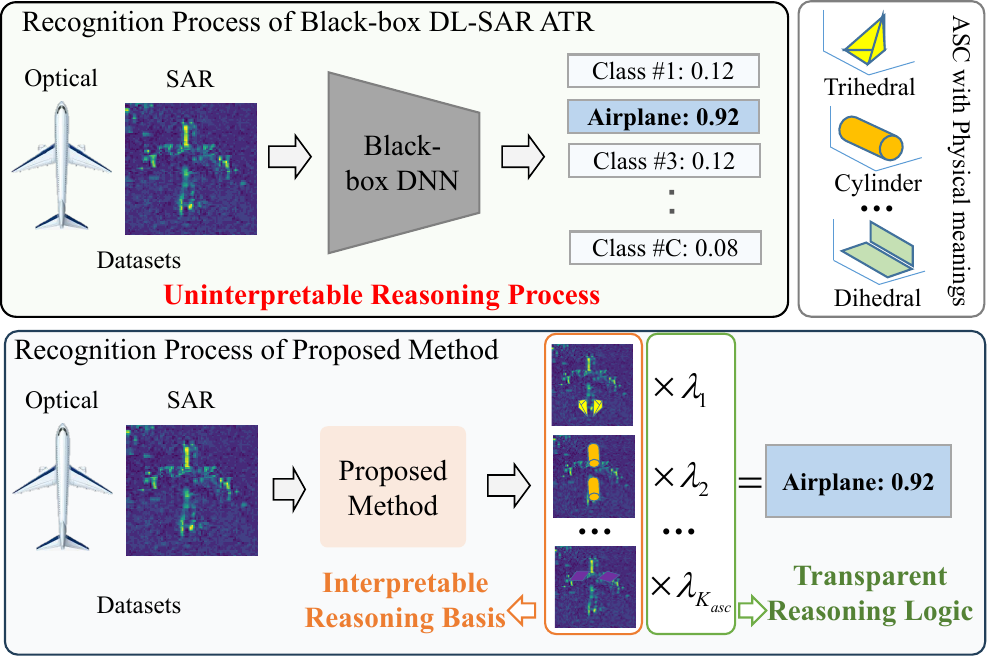}
\caption{Differences between Black-box Deep learning SAR ATR and Proposed Method. The sub-figure above illustrates a classical black-box DL SAR ATR method, where the deep features are uninterpretable and the reasoning logic of the classifier is untransparent. The sub-figure below depicts the proposed method, whose features are physics-based and interpretable, with transparent classifier logic.}
\label{black_grey_DL}
\end{figure}

However, their black-box nature results in low confidence and high risks in practice, which is often unacceptable for decision-critical applications \cite{explain2016proto, rudin2019we, wang2022global}. Due to the lack of a clear and transparent reasoning process, the decisions made by DL-SAR ATR methods are difficult to interpret, hindering their stability and reliability across various scenarios \cite{isprs3, isprs1,wang2022semi}. Despite their superior performance, these challenges limit the practical use of DL-SAR ATR methods.

Therefore, the question arises: can a method be designed that achieves accurate recognition while also providing a clear and transparent reasoning process? In SAR ATR, a transparent reasoning process primarily consists of two key components \cite{rudin2022interpretable}: interpretable features as the reasoning basis and a transparent classifier as the reasoning logic. 
For example, $0.35 \times$ red color + $0.40 \times$ round shape + $0.25 \times$ small size = apple, where features like color, shape, and size serve as the reasoning basis, and the weights (0.35, 0.40, and 0.25) represent the reasoning logic. 
In SAR ATR, the key to an interpretable reasoning process lies in developing interpretable features as the reasoning basis.

In the field of SAR ATR, developing interpretable features as a reasoning basis has garnered significant attention and become a key research focus \cite{XAI_module1}. These approaches typically involve inputting both SAR images and interpretable features (e.g., attribute scattering center-based features) into a deep learning model \cite{XAI_input2}. However, these methods only provide interpretable inputs, while the intermediate deep features used for recognition remain uninterpretable, limiting the transparency of the reasoning process.

In deep learning, two main approaches have been proposed: post-hoc interpretability analysis and importance-based methods \cite{explain1998saliency, explain2010permutation}. Post-hoc analysis methods attempt to reveal the reasoning process of a deep neural network after parameter updates, using techniques like visualization analysis \cite{explain2016cam}, sub-network analysis \cite{subnetwork_2}, or perturbation analysis \cite{explain2010permutation, sensitive_interpret}. However, these methods do not offer intrinsic interpretability and often fail to meet the practical interpretability needs of SAR images \cite{zeiler2014visualizing}. 
Importance-based methods aim to identify key image regions, prototypical cases of classes, or attention maps used by the model for reasoning \cite{explain2016proto, attention_interpret}, but they often do not fully incorporate the physical properties inherent to the SAR domain. Consequently, current deep learning-based SAR ATR methods still lack comprehensive interpretability aligned with SAR-specific features and reasoning processes.

To achieve both accurate recognition and an interpretable reasoning process, we propose an interpretable two-stage feature decomposition method for deep learning-based SAR ATR. This method decomposes uninterpretable deep features into interpretable components with distinct physical meanings. As illustrated in the bottom sub-figure of Fig. \ref{black_grey_DL}, the method provides an interpretable reasoning basis $r_b$ and transparent logic $\lambda_w$, which together form the reasoning logic $\sum_{i} {r_b^i \times \lambda_w^i} = pred$ behind the decisions.

First, ASCCs, as a unique physical parametric model in SAR, can be obtained through clustering methods and offer a concise, physically relevant description of the object \cite{keller1962geometrical}. ASCCs, with their physical geometric types shown in Table \ref{tab:asc_components}, serve as a guide for interpretable features.

However, ASCCs in SAR deep features are often coupled \cite{lecun2015deep, vaswani2017attention}, leading to potential errors in decomposition that undermine interpretability.
This issue highlights the need to optimize deep features based on decomposition errors, which may lack the discriminative power necessary for accurate recognition \cite{ding2018efficient, ac10}. 
To address these challenges, we introduce a feature decoupling and discrimination (FDD) module that pre-decomposes deep features into approximate ASCCs. The FDD module separates deep features into approximate ASCCs with global discriminability.

Next, we propose a multi-layer orthogonal non-negative matrix tri-factorization (MLO-NMTF) approach, which leverages a multi-layer structure to progressively decompose the deep features into linear transformations of multiple ASCCs. The MLO-NMTF model incorporates specific orthogonal constraints, making the entire decomposition mathematically equivalent to a clustering algorithm, which aligns with the method of obtaining ASCCs.

Finally, this method not only achieves accurate recognition but also provides a transparent reasoning process for the decisions, as shown in Fig. \ref{framework}. This approach enhances the trustworthiness of SAR ATR methods, reducing potential risks and improving their practical applicability.

It is important to highlight that our method significantly differs from existing ASC-based SAR ATR approaches. Traditional ASC-based methods primarily treat ASC as the input to a DNN, which struggles to address the inherent black-box nature and lack of interpretability associated with DNNs. In contrast, our method transforms intermediate, non-interpretable deep features into interpretable physical components, thus providing a transparent reasoning process that opens the black box of DNNs.
The key innovations of our method are summarized as follows:

1. Interpretable Features and Transparent Logic: Our method aims to provide interpretable features as the reasoning basis and a transparent classifier to form the reasoning logic behind accurate recognition. This is achieved by decomposing uninterpretable intermediate deep features into interpretable physical components.

2. Two-Stage Feature Decomposition: The FDD module and MLO-NMTF module collaborate to form a two-stage feature decomposition framework. This design ensures that the features used for recognition incorporate both physical geometric properties and discriminative capabilities, leading to accurate recognition while enhancing interpretability.

3. The rationality and effectiveness of our method are validated through extensive experiments on four benchmark datasets. Our method consistently demonstrates superior recognition performance across various network architectures and experimental settings, confirming its strong generalization capability and practical application potential.

The remainder of this paper is organized as follows.
The related work of this paper is presented in Section \ref{Related Work}.
The proposed physics-guided interpretable method is presented in Section \ref{sec: Method}. 
Section \ref{Experiments} validates the rationality and effectiveness of the proposed method.
The conclusions are drawn in Section \ref{conclusions}.

\section{Related Work}
\label{Related Work}

\subsection{SAR ATR Methods}

SAR ATR has evolved significantly, with methods falling into three categories: template-matching-based \cite{temp2016sparse, temp2019sparse,wang2020deep}, model-based \cite{model2011gsc,model2018sca,wang2022sar}, and deep-learning-based approaches \cite{isprs3,deep_hierachical_fewshot, wang2021multiview,wang2023crucial}. The pipeline for deep-learning-based SAR ATR typically involves the following steps:

\begin{equation}
{F_{dl}}\left( {I; Y} \right) = {F_{reg}}\left( {X; Y} \right) \circ {F_{fea}}\left( {I; X} \right)
\end{equation}
where $\circ$ means function composition, $I$ represents the input SAR images, $X$ denotes the corresponding deep features extracted from $I$, and $Y$ is the final decision of the model. The feature extraction process $F_{fea}(X | I) $ often lacks interpretability, while the decision-making process $ F_{reg}(Y | X) $ can be made more interpretable by using a shallow deep model to highlight the weights of components within $ X $.

These data-driven methods have excelled in various SAR ATR tasks such as multi-view recognition, few-shot recognition, and open-set recognition. However, the black-box nature of deep learning models limits interpretability, making it challenging to understand the reasoning behind predictions. This lack of transparency can reduce the confidence in applying these methods in practical SAR ATR scenarios.

\subsection{Interpretable Deep SAR ATR Methods}

Despite the impressive performance of deep learning-based SAR ATR methods, there is still considerable need for interpretable approaches. Interpretability is key to boosting confidence and reliability in practical applications.

Current interpretable SAR ATR methods are mainly input-based approaches, which can be formulated as:
\begin{equation}
{F_{dl}}\left( {\left[ {I,{I_{itp}}} \right]; Y} \right) = {F_{reg}}\left( {X; Y} \right) \circ {F_{fea}}\left( {\left[ {I,{I_{itp}}} \right]; X} \right)
\end{equation}

where $ I_{itp} $ represents the interpretable input. These methods use deep neural networks to extract features from both SAR images and interpretable inputs (e.g., ASC features), which are fused either at the feature or decision level for SAR ATR \cite{XAI_module1, XAI_module2, ac10,wang2023sar,wang2022recognition}. 

While progress has been made in enhancing interpretability, further investigation is required to fully understand the reasoning and logic behind deep learning-based SAR ATR methods. This includes ensuring the physical relevance of the decision-making process and aligning the reasoning with human cognition.

To address these challenges, this paper proposes a physics-guided interpretable method that integrates physics-based reasoning with a transparent decision-making process, represented as:
\begin{equation}
{F_{ours}}\left( {I; Y} \right) = {F_{reg}}\left( {X_{itp};Y} \right) \circ {G_{ours}}\left( {X;X_{itp}} \right) \circ {F_{fea}}\left( {I;X} \right)
\end{equation}
where ${G_{ours}}\left( {{X_{itp}}|X} \right)$ refers to the process in which our method transforms the uninterpretable deep features $X$ into interpretable features ${{X_{itp}}}$. With transparent logic of shadow classifier ${F_{reg}}\left( {Y|X_{itp}} \right)$, our method can provide interpretable features as reasoning basis and achieve precise recognition simultaneously.

\begin{figure*}[tb]
\centering
\includegraphics[width=0.95\textwidth]{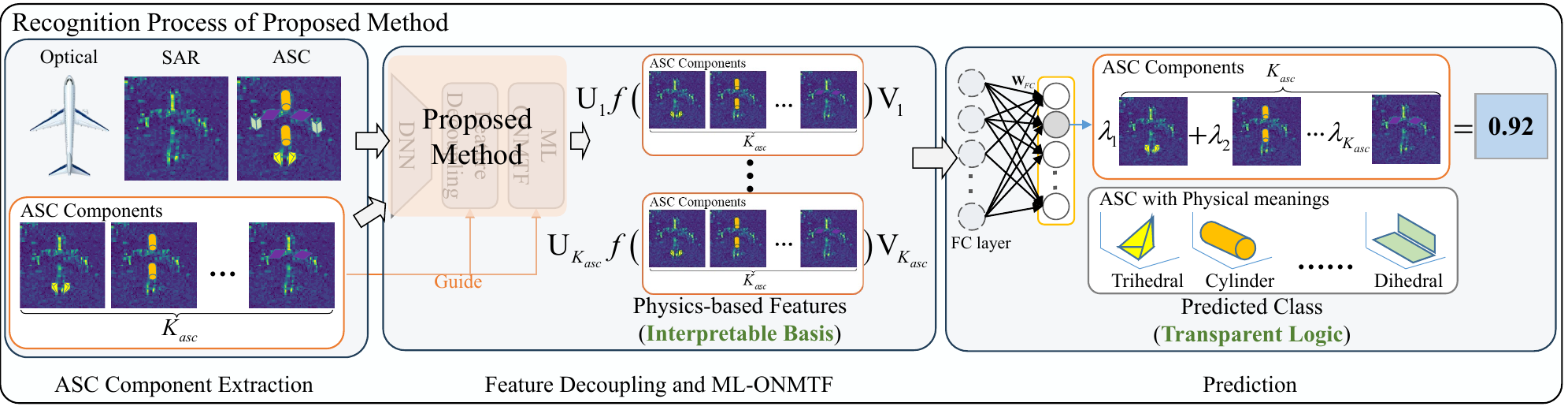}
\caption{The overall framework of the proposed method. First, it employed clustering to obtain ASCCs, followed by a two-stage deep feature decomposition to derive an interpretable reasoning basis. Combined with the transparent reasoning logic of a shallow classifier, the proposed method provides both accurate recognition and an interpretable reasoning process.}
\label{framework}
\end{figure*}

\begin{table}
  \centering
  \caption{Geometric Scattering Types Inference from Frequency Dependence and Length}
  \resizebox{\linewidth}{!}{%
    \begin{tabular}{c|c|c|c}
      \toprule \toprule
      \begin{tabular}[c]{@{}c@{}}Geometric\\scattering type\end{tabular} & \begin{tabular}[c]{@{}c@{}}Frequency\\dependence($\alpha$)\end{tabular} & Length(L) & \begin{tabular}[c]{@{}c@{}}Iconic\\representation\end{tabular}               \\
      \midrule
      Dihedral                                                           & 1                                                                       & $>0$      & \includegraphics[height=0.3cm, keepaspectratio]{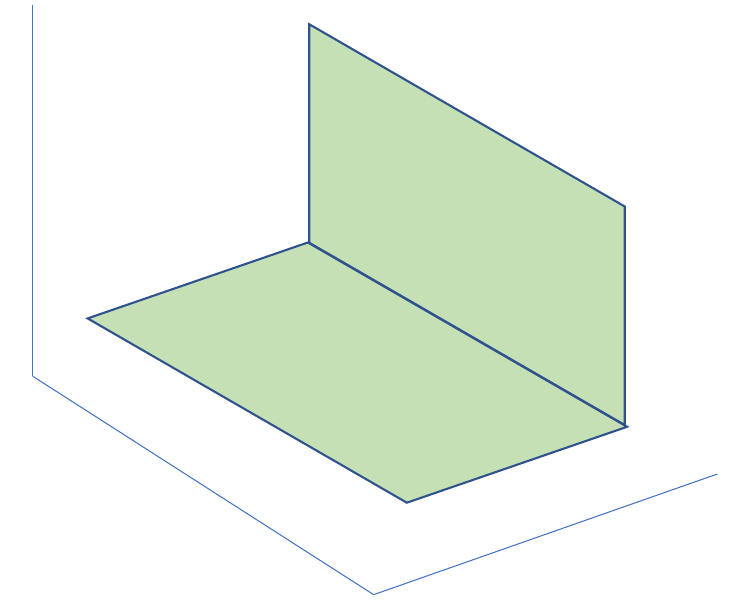}       \\
      \midrule
      Trihedral                                                          & 1                                                                       & 0         & \includegraphics[height=0.3cm, keepaspectratio]{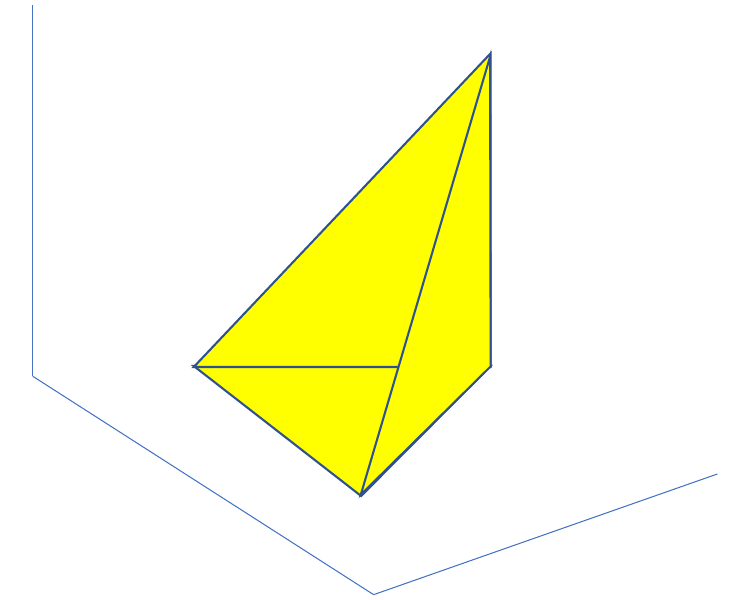}      \\
      \midrule
      Cylinder                                                           & 0.5                                                                     & $>0$      & \includegraphics[height=0.3cm, keepaspectratio]{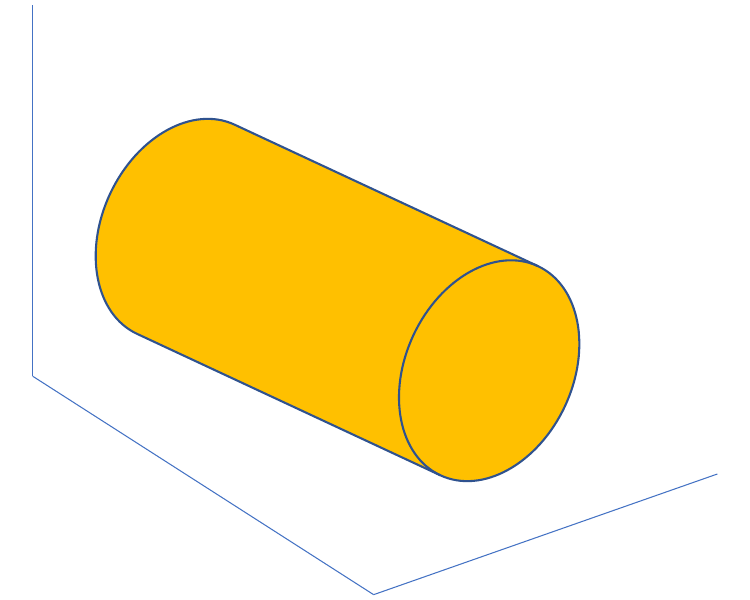}       \\
      \midrule
      Top Hat                                                            & 0.5                                                                     & 0         & \includegraphics[height=0.3cm, keepaspectratio]{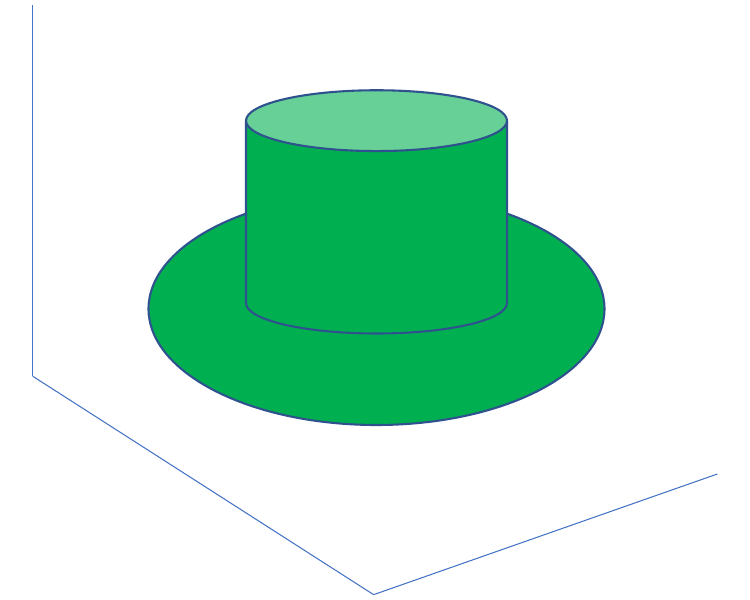}        \\
      \midrule
      Sphere                                                             & 0                                                                       & 0         & \includegraphics[height=0.3cm, keepaspectratio]{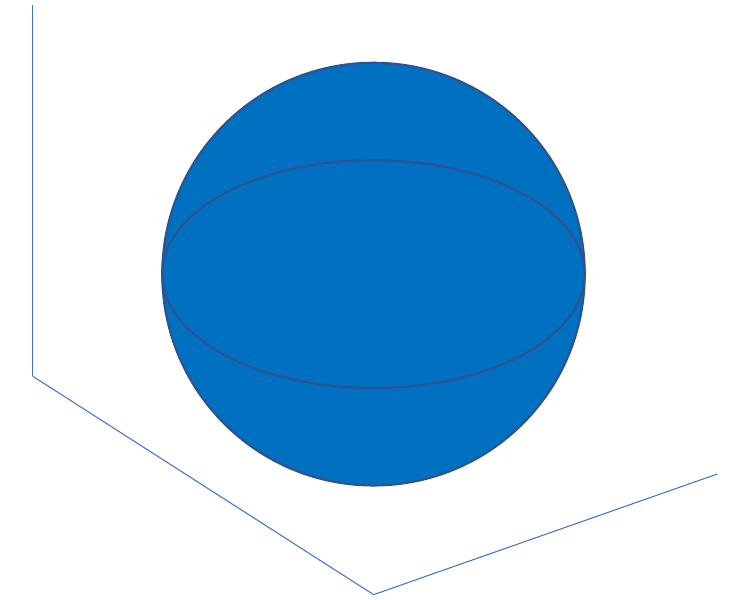}         \\
      \midrule
      Edge Broadside                                                     & 0                                                                       & $>0$      & \includegraphics[height=0.3cm, keepaspectratio]{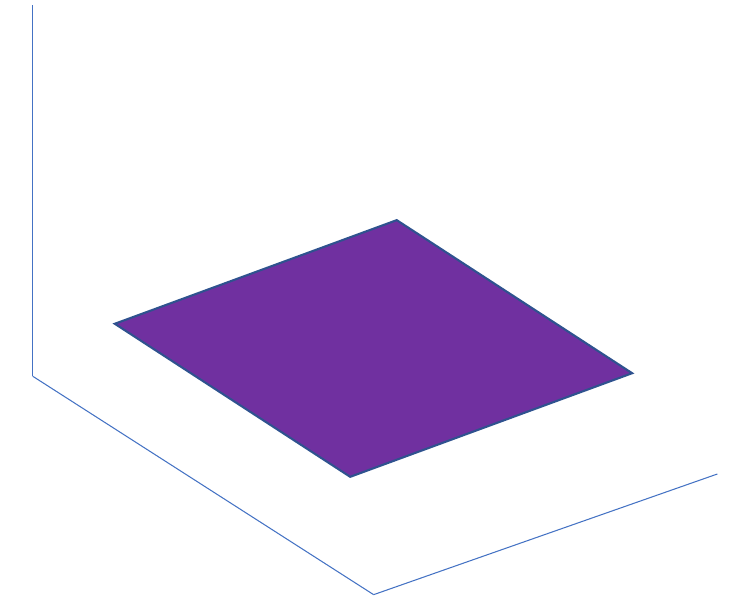} \\
      \midrule
      Edge Diffraction                                                   & -0.5                                                                    & $>0$      & \includegraphics[height=0.3cm]{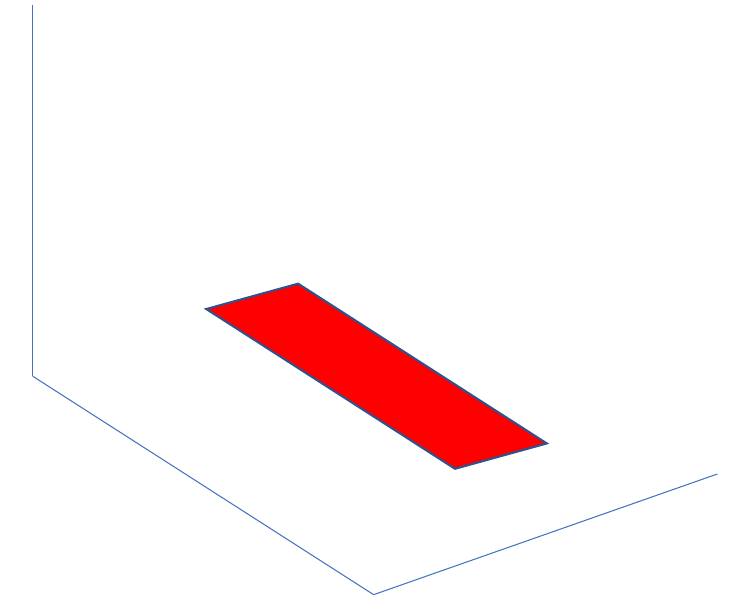}                \\
      \midrule
      Corner Diffraction                                                 & -1                                                                      & 0         & \includegraphics[height=0.3cm]{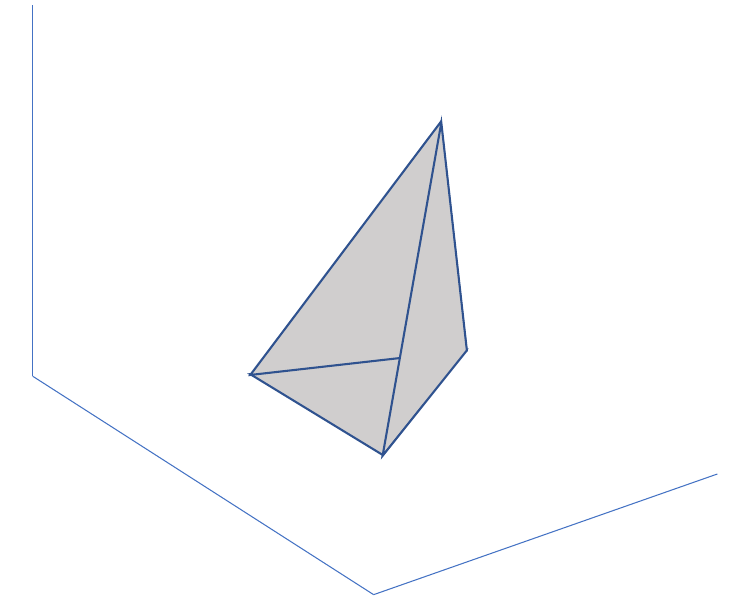}              \\
      \bottomrule \bottomrule
    \end{tabular}
  }
  \label{tab:asc_components}
\end{table}

\section{Proposed Framework}\label{sec: Method}

To construct an interpretable deep ATR method for decision-critical SAR, the key lies in decomposing the uninterpretable deep features into established interpretable bases within the SAR domain. Thus, at first, the proposed interpretable two-stage feature decomposition is designed to extract ASCs and cluster them into ASC components (ASCCs) with different geometric meanings. These ASCCs can serve as the guide for the interpretable reasoning basis, as illustrated in Fig. \ref{framework}.

Next, a feature decoupling and discrimination (FDD) module is introduced to address the conflict between coupled deep features and independent ASCCs with distinct geometric meanings, reducing subsequent decomposition errors. The FDD module also enhances intra-class consistency and inter-class separability of the decomposed features, to preserve accurate recognition performance.

Finally, a multi-layer orthogonal non-negative matrix tri-factorization (MLO-NMTF) is designed and derived to decompose the deep features into multiple ASCCs, in a manner equivalent to a clustering algorithm for obtaining ASCCs. The details of the entire method and all three steps are described as follows.

\subsection{SAR ASC Component Extraction}
\label{method_asc}

At high frequencies, an object's scattering response can be approximated as a sum of responses from individual ASCCs \cite{gerry1999parametric}. 
As an unique physical parametric models of SAR domain, ASCs offer a concise and physically relevant description of the object, capturing both local scattering characteristics and global structure information. 
As shown in Table \ref{tab:asc_components}, different parameter sets of ASCs correspond to distinct geometric meanings. 
For instance, an ASC with frequency $\alpha = 1$ and length $L = 0$ represents a trihedral. 

\begin{figure*}[tb]
\centering
\includegraphics[width=0.95\textwidth]{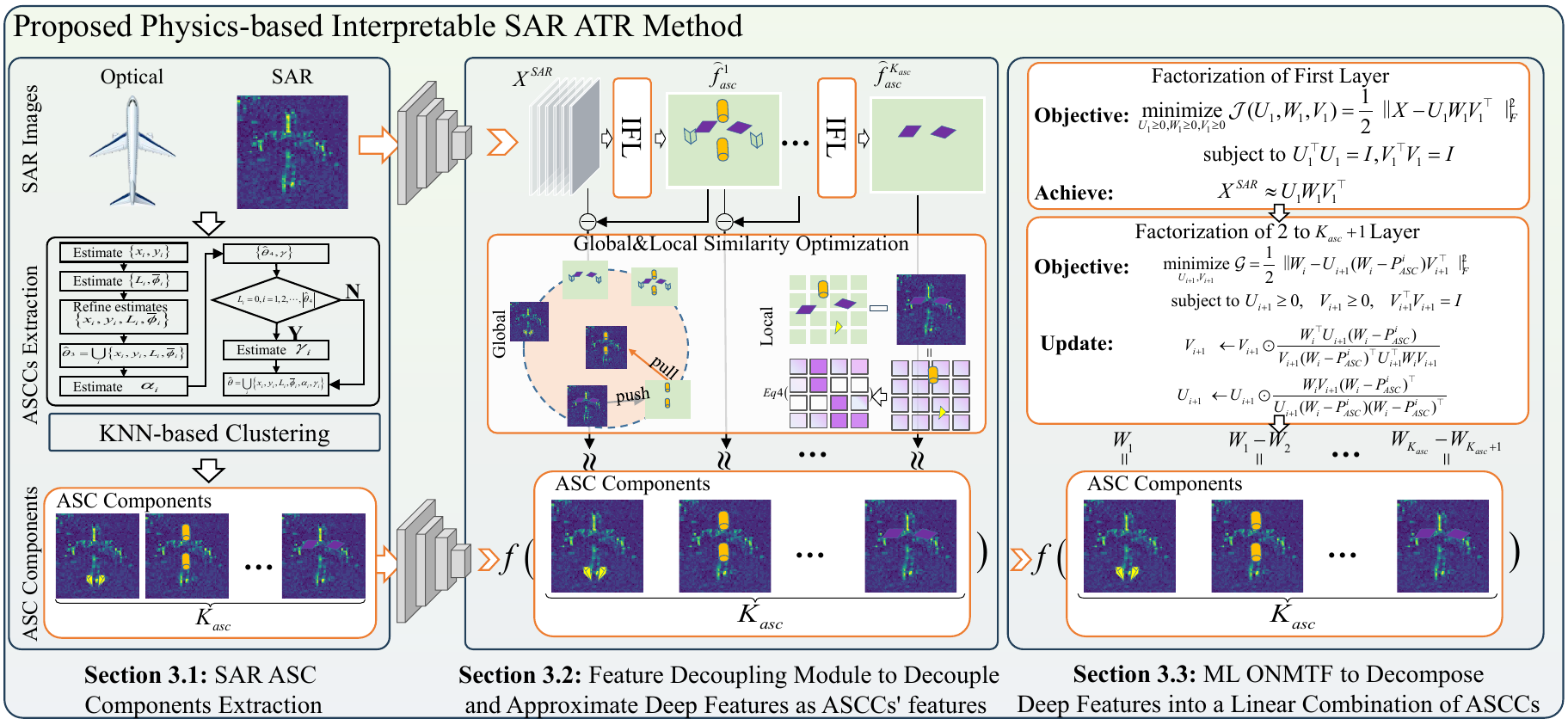}
\caption{Details of the proposed physics-guided two-stage feature decomposition for interpretable deep SAR ATR. }
\label{whole_framework}
\end{figure*}

In this paper, the ASCs guide the transformation of uninterpretable deep features into ASCCs' features with distinct geometric meanings. As shown in Fig. \ref{whole_framework}, each ASC component possesses unique physical characteristics (see Table \ref{tab:asc_components}). The SAR ASC component extraction process is designed in two stages as follows.

Suppose the parameter set $\Theta  = \left\{ {A,x^{pos},y^{pos},\alpha ,L,\bar{\varphi} ,\gamma } \right\}$ includes the amplitude of echoes $A$, relative position $(x^{pos}, y^{pos})$, frequency dependence $\alpha$, length $L$, orientation $\bar{\varphi}$, and aspect dependency $\gamma$. Based on the GTD, the 2-D scattering model is given by \cite{gerry1999parametric}:

\begin{equation}\label{equ:asc1}
\begin{aligned}
E\left( f,\varphi ;\Theta \right) &= \sum\limits_{i = 1}^{K_0} A_i \cdot \left( j\frac{f}{f_c} \right)^{\alpha_i} \cdot \exp \left( -2\pi \gamma_i f \sin \varphi \right) \\
&\quad \cdot \exp \left[j4\pi \frac{f}{v} \left( x_i^{pos} \cos \varphi + y_i^{pos} \sin \varphi \right) \right] \\
&\quad \cdot \mathrm{sinc} \left( 2\pi \frac{L_i}{v} f \sin \left( \varphi - \bar{\varphi}_i \right) \right)
\end{aligned}
\end{equation}
where $f$ is the frequency, $\varphi$ is the aspect angle, $f_c$ is the center frequency, $v$ is the propagation velocity of the electromagnetic wave, and $K_0$ is the number of scattering centers. The estimation of ASCs can be solved using classical OMP methods \cite{liu2017attributed} as well as the latest unfolding net-based methods \cite{liao2024emi}. Following \cite{huang2024physics}, this paper employs the OMP algorithm for estimation:

\begin{equation}\label{equ:asc2}
\mathord{\buildrel{\lower3pt\hbox{$\scriptscriptstyle\frown$}} 
\over \Theta }  = \arg \mathop {\min }\limits_\Theta  {\left\| {S\left( {f,\varphi } \right) - E\left( {f,\varphi ;\Theta } \right)} \right\|_2}
\end{equation}
where ${S\left( {f,\varphi } \right)}$ is the signal in the frequency domain. 

Next, the ASCs are clustered based on the parameter set using K-means clustering. Two clustering methods are used in this paper: the first strictly follows the classification in Table \ref{tab:asc_components}, and clusters the data into 8 ASCCs at most, while the second clusters into $K_{asc}$ ASCCs based on the parameter set. 
For simplicity and clarity, the numbers of ASCCs of clustering are denoted as $K_{asc}$.
Subsequently, the physical parameters of each ASCCs are reconstructed into images, forming multiple images of these ASCCs. For simplicity, in subsequent descriptions, the images of ASCCs will be referred to simply as ASCCs.

Through the above two steps, these ASCCs with different geometric meanings are extracted to serve as the guide for the interpretable basis. 
In the next subsection, the FDD module is proposed to pre-decompose deep features and improve the discriminability of post-decomposition features, to reduce MLO-NMTF's errors and ensure accurate recognition.

\subsection{Feature Decoupling and Discrimination Module} \label{methodsec1}

After obtaining ASCCs with clear geometric meanings in the previous section, the key challenge now is to transform the inherently uninterpretable deep features into interpretable reasoning basis. Our solution is to decompose these deep features into interpretable ASCCs, leveraging the fact that SAR images can be approximated as a linear transformation of ASCs \cite{keller1962geometrical}.

However, due to the nature of DNN operations when extracting features \cite{lecun2015deep, zeiler2014visualizing, vaswani2017attention, wang2025limited}, the scatter points on SAR images become coupled in the deep features. On the other hand, 
components with distinct physical meanings are relatively independent \cite{wang2012nonnegative, ding2005equivalence,li2025volume}, which means that the deep features need to be optimized based on decomposition errors. However, optimizing deep features solely based on decomposition errors is similar to directly using ASCCs for recognition, which may lack the discriminative power needed for precise recognition \cite{ding2018efficient, ac10}.

To address this issue, the FDD module is designed to pre-decompose deep features into approximate ASCCs while enhancing the discriminability of both pre- and post-decomposition features. The FDD module reduces subsequent decomposition errors while maintaining sufficient discriminative power for precise recognition, as shown in Fig. \ref{whole_framework}. It operates on two levels: Globally, it ensures that both pre- and post-decomposition features exhibit inter-class separation and intra-class compactness, while locally, it ensures that the decomposed approximate ASCCs closely match the true ASCCs at each pixel. 

The pipeline of the feature decoupling module is described as follows:

Given an SAR image, and the corresponding ASCCs, $\left\{ {{\mathbf{x}}_{i1}^{asc},{\mathbf{x}}_{i2}^{asc}, \ldots ,{\mathbf{x}}_{i{K_{asc}}}^{asc}} \right\}$. The SAR images and ASCCs first go through any DNN to obtain the deep features ${X^{SAR}}$ and the ASCCs’ features $\left\{ {P_{ASC}^{1}, P_{ASC}^{2}, \ldots ,P_{ASC}^{K_{asc}}} \right\} \in {\mathbb{R}^{h \times w \times c}}$.

Then, the feature decoupling module utilizes $K_{asc}$ information filter layers to incrementally filter out $K_{asc}$ intermediate features $f_{ij}^{mid}$ from ${X^{SAR}}$. This is achieved through global and local similarity constraints, ensuring:
\begin{align}
{X^{SAR}} - f_{i1}^{mid}: = \overset{\hbox{$\smash{\scriptscriptstyle\frown}$}}{P}^{1}_{ASC} \approx P_{ASC}^{1} \\
f_{i-1}^{mid} - f_{i}^{mid}: = \overset{\hbox{$\smash{\scriptscriptstyle\frown}$}}{P}_{ASC}^{i} \approx P_{ASC}^{i}
\end{align}

Globally, a similarity constraint is applied by calculating the similarity between $\overset{\hbox{$\smash{\scriptscriptstyle\frown}$}}{P}_{ASC}^{i}$ and $P_{ASC}^{i}$, optimizing $\overset{\hbox{$\smash{\scriptscriptstyle\frown}$}}{P}_{ASC}^{i}$ to closely approximate $P_{ASC}^{i}$ with enhanced inter-class separation and intra-class compactness:

\begin{align}
    Ld_{ik_{n}}= -\sum_k^{K} \log \frac{{{d^t(f_{ik},P_i)}}}{{{d^t(f_{ik},P_i)}} + \sum_{f_{jk}\in D_{{\lnot A}k_{n}}} {d^t(f_{jl},P_i)}}
\end{align}
where $d^t(\cdot)$ represents cosine similarity for scale invariance \cite{cos4}. 

Locally, pixels with large errors are filtered out using a threshold and weighted more heavily to ensure $\overset{\hbox{$\smash{\scriptscriptstyle\frown}$}}{f}_{ij}^{asc}$ approximates $f_{ij}^{asc}$ at each pixel:
\begin{equation}
L_p^i= - \sum_k^K \lambda_{iw}^k *d^t\left (l2n\left (f_{ik}\right ),l2n\left (P_i\right )\right )
\end{equation}
where $l2n(\cdot)$ denotes L2 normalization. The weight $\lambda_{iw}^k$ is computed as follows:
\begin{equation}
\lambda_{iw}^k = \left (1-\beta * \frac{d^t\left (l2n\left (f_{ik}\right ),l2n\left (P_i\right )\right )+2}{2} \right )^{\rho }
\end{equation}
where $\rho \ge 0$ is a parameter that adjusts $\lambda_{iw}^k$, and $\beta = 1$ if $\frac{d^t\left ( l2n\left (f_{ik}\right ), l2n\left (P_i\right )\right ) - d^{t-1}\left ( l2n\left (f_{ik}\right ), l2n\left (P_i\right )\right )}{d^t\left ( l2n\left (f_{ik}\right ), l2n\left (P_i\right )\right )} \ge \epsilon$, otherwise $\beta = 0$. This serves as a gate. $\lambda_{iw}^k$  corrects the weights of pixels with large errors.

Through the process described above, the FDD module can reduce the error in subsequent decomposition while ensuring that the decomposed features maintain discriminability for accurate recognition.

\subsection{Multi-Layer Orthogonal NMTF} \label{methodsec2}

The multi-layer orthogonal NMTF (MLO-NMTF) module in this subsection primarily aims to decompose deep features into multiple ASCCs with clear geometric meanings. Since SAR images can be approximated as a linear transformation of ASCCs \cite{keller1962geometrical}, the MLO-NMTF module uses a multi-layer structure to gradually decompose the features into distinct ASCCs. By incorporating specific orthogonal constraints in the tri-factorization form, the module becomes mathematically equivalent to a clustering algorithm \cite{yoo2010orthogonal}, which aligns with the method of obtaining ASCCs.

This MLO-NMTF module employs multiplicative updates on the Stiefel manifold, directly utilizing true gradient information while maintaining orthogonality constraints.
The module is divided into two main steps:

(1) factorization of the first layer:
For simplicity and clarity, $X \in \mathbb{R}^{h \times w}$ will represent a single matrix along the channel dimension of $X^{SAR}$ throughout the following contents. 
Note that the deep features $X \in \mathbb{R}^{m \times n}$ can be transformed to be non-negative. 
To find the low-rank approximation of $X$, it is decomposed into three matrices with two orthogonal constraints \cite{choi2008algorithms}:
\begin{equation}\label{equ: X=U1W1V1}
    X\approx U_{1}W_{1}V_{1}^{\top},
\end{equation}
where $U_{1} \in \mathbb{R}^{h \times r}$ and $V_{1} \in \mathbb{R}^{w \times r}$ are orthogonal matrices satisfying $U_{1}^{\top}U_{1} = I$ and $V_{1}^{\top}V_{1}=I$, $r$ is the target low rank. 
The two orthogonal constraints not only ensure that the matrix $W_{1} \in \mathbb{R}^{r \times r}$ is a low-rank approximation of deep features $X$, but also make this decomposition essentially equivalent to clustering \cite{ding2005equivalence}, which is well-suited for the acquisition of ASCCs.

The factorization above can be transformed into the following optimization problem:
\begin{equation}\label{equ: J = X-U1W1V1}
    \begin{aligned}
    & \underset{U_{1}\geq 0,W_{1}\geq 0,V_{1}\geq 0}{\text{minimize}}
    & & \mathcal{J}(U_{1},W_{1},V_{1}) = \frac{1}{2}\|X-U_{1}W_{1}V_{1}^{\top}\|_{F}^{2}  \\
    & \text{subject to}
    & & U_{1}^{\top}U_{1} = I , V_{1}^{\top}V_{1}=I.
\end{aligned}
\end{equation}

Multiplicative updates are employed to directly utilize true gradient information for the optimization problem.
Suppose that the gradient of an optimization function $\mathcal{J}$ has a decomposition that is of the form,
\begin{equation}\label{equ: J = J+ - J-}
    \nabla \mathcal{J} = [\nabla \mathcal{J}]^{+}-[\nabla \mathcal{J}]^{-},
\end{equation}
where $[\nabla \mathcal{J}]^{+}>0$ and $[\nabla \mathcal{J}]^{-}>0$. 
Then multiplicative update for parameters $\Theta$ has the form
\begin{equation}\label{equ: Theta <- Theta}
    \Theta \leftarrow \Theta \odot \left(\frac{[\nabla \mathcal{J}]^{-}}{[\nabla \mathcal{J}]^{+}}\right)^{\eta},
\end{equation}
where $\odot$ represents Hadamard product and $(\centerdot)^{\eta}$ denotes the elementwise power and $\eta (0<\eta \leq 1)$ is a learning rate. 
It can be easily seen that the multiplicative update \eqref{equ: J = X-U1W1V1} preserves the nonnegativity of the parameter $\Theta$, while $\nabla \mathcal{J} = 0$ when the convergence is achieved.

Based on these gradient calculations, rule \eqref{equ: Theta <- Theta} with $\eta = 1$ yields the multiplicative updates on Stiefel manifolds \cite{stiefel1935richtungsfelder} by
\begin{align}
    U_{1} &\leftarrow U_{1} \odot \frac{XV_{1}W_{1}^{\top}}{U_{1}W_{1}V_{1}^{\top}X^{\top}U_{1}}, \label{equ: update U1}\\
    V_{1} &\leftarrow V_{1}\odot\frac{X^{\top}U_{1}W_{1}}{V_{1}W_{1}^{\top}U_{1}^{\top}XV_{1}},\label{equ: update V1}\\
    W_{1} &\leftarrow W_{1}\odot\frac{U_{1}^{\top}XV_{1}}{U_{1}^{\top}U_{1}W_{1}V_{1}^{\top}V_{1}}.\label{equ: update W1}
\end{align}

(2) Factorization of the $2$ to $K_{asc}+1$ layer:
Once the low-rank principal components $W_{i}$ are extracted from $X$, an additional equality constraint related to ASCCs is imposed during the decomposition of $W_{i} \in \mathbb{R}^{r \times r}$. 
Thus, through multi-layer decomposition, the data matrix is then decomposed into ASCCs:
\begin{equation}
    W_{i}\approx U_{i+1}W_{i+1}V_{i+1}^{\top}, \quad W_{i}-W_{i+1}=P_{ASC}^{i}, 
\end{equation}
where the condition $W_{i}-W_{i+1}=P_{ASC}^{i}$ implies that the ASC feature in the image is a superposition of layers and $U_{i+1}, V_{i+1}\in \mathbb{R}^{r \times r}$.

To make the factorization of the $2$ to $K_{asc}+1$ layer be equivalent to a k-means clustering used in ASCCs extraction, according to the theory of ONMTF \cite{ding2005equivalence}, the factorization is transformed into 
\begin{equation}\label{equ: orthogonal non-negative matrix factorization with constraints problem}
\begin{aligned}
& \underset{U_{i+1}\geq 0,V_{i+1}\geq 0}{\text{minimize}}
& & \mathcal{G}(U_{i+1},V_{i+1}) = \frac{1}{2}\|W_{i}-U_{i+1}W_{i+1}V_{i+1}^{\top}\|_{F}^{2}  \\
& \text{subject to}
& & W_{i}-W_{i+1}=P_{ASC}^{i}, \quad V_{i+1}^{\top}V_{i+1}=I.
\end{aligned}
\end{equation}
By substituting $W_{i}-W_{i+1}=P_{ASC}^{i}$ from the constraint into the optimization problem, \eqref{equ: orthogonal non-negative matrix factorization with constraints problem} can be transformed into the following problem,
\begin{equation}\label{equ: orthogonal non-negative matrix factorization problem}
\begin{aligned}
& \underset{U_{i+1},V_{i+1}}{\text{minimize}}
& & \mathcal{G} = \frac{1}{2}\|W_{i}-U_{i+1}(W_{i}-P_{ASC}^{i})V_{i+1}^{\top}\|_{F}^{2}  \\
& \text{subject to}
& &  U_{i+1}\geq 0,\quad V_{i+1}\geq 0, \quad V_{i+1}^{\top}V_{i+1}=I.
\end{aligned}
\end{equation}
To solve this optimization problem \eqref{equ: orthogonal non-negative matrix factorization problem}, the multiplicative update strategy from problem \eqref{equ: J = X-U1W1V1} is employed here again.
First, $U_{i+1}$ is considered, whose gradient in $\mathcal{G}$ is given by
\begin{equation}\label{equ: nabla Ui G}
    \begin{aligned}
        \nabla_{U_{i+1}} \mathcal{G} & = - W_{i}V_{i+1}(W_{i}-P_{ASC}^{i})^{\top}  \\
        & + U_{i+1}(W_{i}-P_{ASC}^{i})V_{i+1}^{\top}V_{i+1}(W_{i}-P_{ASC}^{i})^{\top}\\
        & = - W_{i}V_{i+1}(W_{i}-P_{ASC}^{i})^{\top} \\
        & +U_{i+1}(W_{i}-P_{ASC}^{i})(W_{i}-P_{ASC}^{i})^{\top}.
    \end{aligned}
\end{equation}
Note that $V_{i+1}$ is non-negative and orthogonal.
The orthogonality constraint on $V_{i+1}$ is a restriction on the Stiefel manifold \cite{stiefel1935richtungsfelder}.
The gradient on the Stiefel manifold can be calculated using
\begin{equation}\label{equ: nabla Vi G = nabla Vi G - Vi nabla Vi G Vi}
    \widetilde{\nabla}_{V_{i+1}} \mathcal{G} = \nabla_{V_{i+1}} \mathcal{G} - V_{i+1} (\nabla_{V_{i+1}} \mathcal{G})^{\top}V_{i+1},
\end{equation}
where the gradient $\nabla_{V_{i+1}} \mathcal{G}$ can be computed as
\begin{equation}\label{equ: nabla Vi G = -WU(W-P) +V(W-P)UU(W-P)}
    \begin{aligned}
        \nabla_{V_{i+1}} \mathcal{G} & = -W_{i}^{\top}U_{i+1}(W_{i}-P_{ASC}^{i}) \\
        &+V_{i+1}(W_{i}-P_{ASC}^{i})^{\top}U_{i+1}^{\top}U_{i+1}(W_{i}-P_{ASC}^{i}).
    \end{aligned}
\end{equation}

Substituting \eqref{equ: nabla Vi G = -WU(W-P) +V(W-P)UU(W-P)} into \eqref{equ: nabla Vi G = nabla Vi G - Vi nabla Vi G Vi} yields the gradient on the Stiefel manifold,
\begin{equation}\label{equ: nabla V G in Stiefel manifold}
    \begin{aligned}
        \widetilde{\nabla}_{V_{i+1}} \mathcal{G} & = - W_{i}^{\top}U_{i+1}(W_{i}-P_{ASC}^{i}) \\
        & + V_{i+1}(W_{i}-P_{ASC}^{i})^{\top}U_{i+1}^{\top}W_{i}V_{i+1}.
    \end{aligned}
\end{equation}

The optimization problem \eqref{equ: orthogonal non-negative matrix factorization problem} can be solved through the following iterative updates,
\begin{equation}\label{equ: Wi+1 Hi+1 update in ONMF}
    \begin{aligned}
        V_{i+1} &\leftarrow V_{i+1}\odot\frac{W_{i}^{\top}U_{i+1}(W_{i}-P_{ASC}^{i})}{V_{i+1}(W_{i}-P_{ASC}^{i})^{\top}U_{i+1}^{\top}W_{i}V_{i+1}},\\
        U_{i+1} &\leftarrow U_{i+1}\odot \frac{W_{i}V_{i+1}(W_{i}-P_{ASC}^{i})^{\top}}{U_{i+1}(W_{i}-P_{ASC}^{i})(W_{i}-P_{ASC}^{i})^{\top}}.\\
    \end{aligned}
\end{equation}

Thus, the decomposition of $X$ into the ASCCs can be obtained by 
\begin{align}
    X \approx U_{1}W_{1}V_{1},&  \label{equ: X U1W1V1} \\
    W_{1}\approx U_{2}W_{2}V_{2}^{\top},& \quad W_{1}-W_{2}=P_{ASC}^{1}, \label{equ: W1=U2W2V2}\\
    W_{2}\approx U_{3}W_{3}V_{3}^{\top},& \quad W_{2}-W_{3}=P_{ASC}^{2}, \label{equ: W2=U3W3V3}\\
    \cdots &\nonumber\\
    W_{k}\approx U_{k+1}W_{k+1}V_{k+1}^{\top},& \quad W_{k+1}-W_{k}=P_{ASC}^{k}. \label{equ: Wk=Uk+1Wk+1Vk+1}
\end{align}

Finally, the MLO-NMTF can decompose the original uninterpretable $X^{SAR}$ into the interpretable ASCCs with clear geometric meanings.  

At this point, a transparent classifier with only one dense layer is employed to directly use the decomposed $W \in \mathbb{R}^{c \times \left( {{K_{asc}} + 1} \right) \times r \times r}$ to perform the final recognition. 
This ensures that the entire framework has a transparent recognition process.
The details of the transparent classifier are shown in Fig. \ref{whole_framework} and the experiment section.

It should be noted that the proposed method is not strictly dependent on the physical geometric properties of ASCCs. Any independent components with clear and meaningful interpretations can replace ASCCs for use within the proposed method.

Through the three subsections described above, the proposed physics-based two-stage feature decomposition not only achieves precise recognition but also provides an interpretable reasoning process behind the decisions. 

\section{Experiments and Results}
\label{Experiments}

This section starts with an introduction to the benchmark datasets, along with preprocessing procedures and training details. It then validates the interpretability and recognition performance of the proposed interpretable SAR ATR method.

First, the method's interpretability is validated in two aspects:
1. Rationality Validation: Evaluating whether the ASCCs exhibit intra-class compactness and inter-class separability, ensuring their effectiveness for recognition.
2. Effectiveness Validation: Assessing the accuracy and reliability of decomposing uninterpretable deep features into a linear combination of ASCC features.

Next, recognition performance is validated through experiments on four benchmark datasets: MSTAR, OpenSAR-Ship, SAR-AIRcraft, and FUSAR-Ship, with varying training sample sizes to evaluate the method’s effectiveness.

Finally, the accuracy of deep feature decomposition and recognition performance under different configurations is validated, including:
1. Different deep learning methods (CNN-based and Transformer-based architectures). 2. Varying numbers of ASCCs.

The recognition performance is also compared with other state-of-the-art SAR ATR methods.

\begin{table}[htb]
\renewcommand{\arraystretch}{1.2}
\setlength\tabcolsep{4.4pt}
\centering
\footnotesize
\caption{Original Image Number of Different Depressions for SOC in MSTAR Dataset}
\label{ttnumMSTAR}
\begin{tabular}{c|cc|cc}
\toprule \toprule
\multirow{2}{*}{Class} & \multicolumn{2}{c|}{Training}            & \multicolumn{2}{c}{Testing}             \\ \cline{2-5} 
                       & \multicolumn{1}{c|}{Number} & Depression & \multicolumn{1}{c|}{Number} & Depression \\ \midrule
BMP2-9563 & \multicolumn{1}{c|}{233} & \multirow{9}{*}{$\text{17}{}^\circ$} & \multicolumn{1}{c|}{195} & \multirow{9}{*}{$\text{15}{}^\circ$} \\ \cline{1-2} \cline{4-4}
BRDM2-E71              & \multicolumn{1}{c|}{298}    &            & \multicolumn{1}{c|}{274}    &            \\ \cline{1-2} \cline{4-4}
BTR60-7532             & \multicolumn{1}{c|}{256}    &            & \multicolumn{1}{c|}{195}    &            \\ \cline{1-2} \cline{4-4}
BTR70-c71              & \multicolumn{1}{c|}{233}    &            & \multicolumn{1}{c|}{196}    &            \\ \cline{1-2} \cline{4-4}
D7-92                  & \multicolumn{1}{c|}{299}    &            & \multicolumn{1}{c|}{274}    &            \\ \cline{1-2} \cline{4-4}
2S1-b01                & \multicolumn{1}{c|}{299}    &            & \multicolumn{1}{c|}{274}    &            \\ \cline{1-2} \cline{4-4}
T62-A51                & \multicolumn{1}{c|}{299}    &            & \multicolumn{1}{c|}{273}    &            \\ \cline{1-2} \cline{4-4}
T72-132                & \multicolumn{1}{c|}{232}    &            & \multicolumn{1}{c|}{196}    &            \\ \cline{1-2} \cline{4-4}
ZIL131-E12             & \multicolumn{1}{c|}{299}    &            & \multicolumn{1}{c|}{274}    &            \\ \cline{1-2} \cline{4-4}
ZSU234-d08             & \multicolumn{1}{c|}{299}    &            & \multicolumn{1}{c|}{274}    &            \\ \bottomrule \bottomrule
\end{tabular}
\end{table}

\renewcommand{\arraystretch}{1.5}
\begin{table}[htbp]
\centering
\scriptsize
\caption{Image Number and Imaging Conditions of Different Targets in OpenSARShip Dataset}
\label{opensarset}
\setlength\tabcolsep{2.2pt}
\begin{tabular}{c|c|ccc}
\toprule \toprule
Class          & Imaging Condition                                                                  & \begin{tabular}[c]{@{}c@{}}Training\\ Number\end{tabular} & \begin{tabular}[c]{@{}c@{}}Testing\\ Number\end{tabular} & \begin{tabular}[c]{@{}c@{}}Total\\ Number\end{tabular} \\ \midrule 
Bulk Carrier   & \multirow{6}{*}{\begin{tabular}[c]{@{}c@{}} VH and VV, C band\\  Resolution=$5-20$m\\ Incident angle=$20^{\circ}-45^{\circ}$ \\ Elevation sweep angle=$\pm 11^{\circ}$\\ ${\text{Rg20}}m \times {\text{az}}22m$\end{tabular}} & 200                                                       & 475                                                      & 675                                                    \\ \cline{1-1} \cline{3-5}
Container Ship &                                                                                    & 200                                                       & 811                                                      & 1011                                                   \\ \cline{1-1} \cline{3-5}
Tanker         &                                                                                    & 200                                                       & 354                                                      & 554                                                    \\ \cline{1-1} \cline{3-5}
Cargo          &                                                                                    & 200                                                       & 557                                                      & 757                                                    \\ \cline{1-1} \cline{3-5}
Fishing        &                                                                                    & 200                                                       & 121                                                      & 321                                                    \\ \cline{1-1} \cline{3-5}
General Cargo  &                                                                                    & 200                                                       & 165                                                      & 365     \\ \bottomrule \bottomrule                                               
\end{tabular}
\end{table}

\renewcommand{\arraystretch}{1.5}
\begin{table}[htbp]
\centering
\scriptsize
\caption{Image Number and Imaging Conditions of Different Targets in FUSAR-Ship Dataset}
\label{fusarset}
\setlength\tabcolsep{1.5pt}
\begin{tabular}{c|c|ccc}
\toprule \toprule 
Class          & Imaging Condition                                                                  & \begin{tabular}[c]{@{}c@{}}Training\\ Number\end{tabular} & \begin{tabular}[c]{@{}c@{}}Testing\\ Number\end{tabular} & \begin{tabular}[c]{@{}c@{}}Total\\ Number\end{tabular} \\ \midrule  
Bulk Carrier          & \multirow{5}{*}{\begin{tabular}[c]{@{}c@{}}VH and VV, C band\\  Resolution=$0.5-500$m\\ Incident angle=$10^{\circ}-60^{\circ}$ \\ Elevation sweep angle=$\pm 20^{\circ}$\\ ${\text{Rg20}}m \times {\text{az}}22m$\end{tabular}} & 100                  & 173                  & 273                  \\ \cline{1-1} \cline{3-5}
Cargo Ship            &                                                                                    & 100                  & 1593                 & 1693                 \\ \cline{1-1} \cline{3-5}
Fishing              &                                                                                    & 100                  & 685                  & 785                  \\ \cline{1-1} \cline{3-5}
Other type of ship      &                                                                                    & 100                  & 1507                 & 1607                 \\ \cline{1-1} \cline{3-5}
Tanker               &                                                                                    & 100                  & 48                   & 148                  \\ \bottomrule \bottomrule                                               
\end{tabular} 
\end{table}

\renewcommand{\arraystretch}{1.5}
\begin{table}[htbp]
    \centering
    \scriptsize
    \caption{Image Number and Imaging Conditions of Different Targets in SAR-Aircraft Dataset}
    \label{saracd}
    \setlength\tabcolsep{2.3pt}
    \begin{tabular}{c|c|ccc}
        \toprule \toprule
        Class                       & Imaging Condition &
        \begin{tabular}[c]{@{}c@{}}
            Training \\ Number
        \end{tabular} &
        \begin{tabular}[c]{@{}c@{}}
            Testing \\ Number
        \end{tabular} &
        \begin{tabular}[c]{@{}c@{}}
            Total \\ Number
        \end{tabular}                                        \\ \midrule
        A220                        & \multirow{6}{*}{
            \begin{tabular}[c]{@{}c@{}}
                Spotlight mode                         \\
                HH polarization                        \\
                Resolution=$1$m                        \\
                Incident angle=$10^{\circ}-60^{\circ}$ \\
                Elevation sweep angle=$\pm 20^{\circ}$
            \end{tabular}
        }
                                                        & 150 & 314 & 464       \\ \cline{1-1} \cline{3-5}
        A330                        &                   & 150 & 362 & 512 \\ \cline{1-1} \cline{3-5}
        A320/321                    &                   & 150 & 360 & 510 \\ \cline{1-1} \cline{3-5}
        ARJ21                       &                   & 150 & 364 & 514 \\ \cline{1-1} \cline{3-5}
        Boeing737                   &                   & 150 & 378 & 528 \\ \cline{1-1} \cline{3-5}
        Boeing787                   &                   & 150 & 354 & 504 \\ \bottomrule \bottomrule
    \end{tabular}
\end{table}

\subsection{Datasets and Configuration}

In this section, we use the MSTAR, OpenSARship, FUSAR-ship, and SAR-AIRcraft datasets to validate the interpretability and recognition performance of the proposed interpretable SAR ATR method.

The MSTAR dataset, released by DARPA and AFRL, serves as a benchmark for SAR ATR performance. It consists of SAR images with 1-foot resolution in the X-band, collected using the STARLOS sensor platform. The dataset includes ten types of ground targets, such as tanks, rocket launchers, and bulldozers, captured at various aspect and depression angles.

The OpenSARship dataset is designed to develop ship detection and classification algorithms under high interference \cite{OpenSARShip}. It consists of 41 Sentinel-1 images and 11,346 ship chips, representing 17 ship types. The ship labels are based on AIS information, with ship lengths ranging from 92m to 399m and widths from 6m to 65m. Both VV and VH data are used for training, validation, and testing.

The FUSAR-Ship dataset is another open benchmark specifically designed for ship and marine target detection and recognition \cite{FUSAR}. Compiled by the Key Lab of Information Science of Electromagnetic Waves (MoE) at Fudan University for the Gaofen-3 satellite, it serves as an open SAR-AIS matching dataset. This dataset consists of over 100 Gaofen-3 scenes and more than 5000 ship image slices with corresponding AIS information. It offers distinct imaging parameters, such as incident angle, bandwidth, and resolution, which have larger ranges compared to the OpenSARship dataset.

The SAR-AIRcraft dataset is obtained from the GF-3 satellite, using Spotlight mode for imaging with a 1m resolution. The dataset covers the Shanghai Hongqiao Airport, Beijing Capital International Airport, and another airport in different time phases with HH polarization \cite{saracd}. It includes 4368 images and 16,463 instances of aircraft targets. The dataset is characterized by complex scenes, diverse target categories, dense target distribution, and noise interference.

The training and network configurations are as follows. The input SAR images are resized to $224 \times 224$ pixels using bilinear interpolation. The value of $K_{asc}$ is set to 6, except otherwise specified. The learning rate is initialized at 0.1 for CNN-based experiments and 0.001 for Transformer-based experiments, with a reduction factor of 0.5 every 15 epochs.
The details of the division of the training and test data for the four benchmark datasets are shown in Table~\ref{ttnumMSTAR}-Table~\ref{saracd} respectively.
The proposed method is tested and evaluated on a GPU cluster equipped with 20 CPUs (Intel(R) Xeon(R) Platinum 8457C) and 12 NVIDIA A100 GPUs with 32GB of memory each. The implementation is carried out using the open-source PyTorch framework on four Tesla A100 GPUs.

\begin{figure}[tb]
\centering
\includegraphics[width=0.48\textwidth]{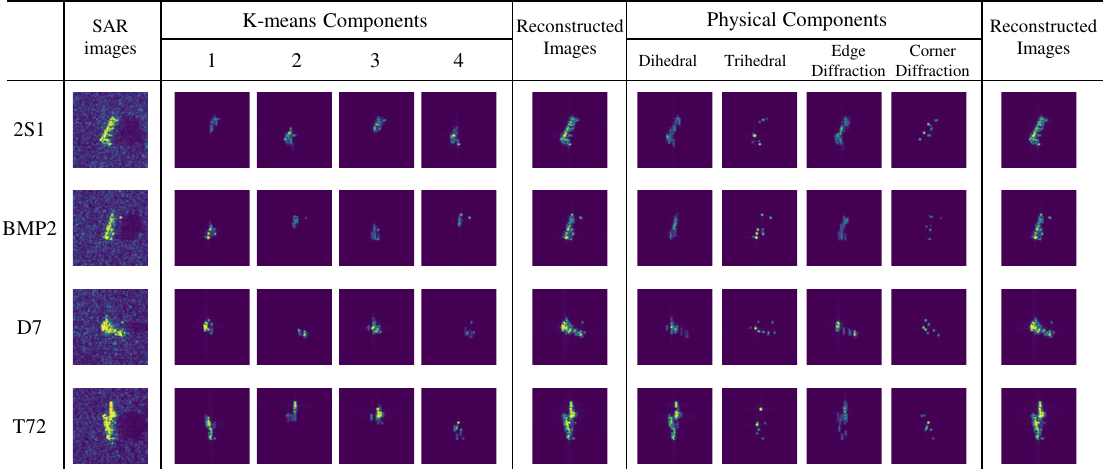}
\caption{ASCCs of different SAR targets and their corresponding reconstructed SAR images under two clustering methods. When clustered based on physical meaning, the different ASCCs exhibit distinct geometric types, like dihedral.}
\label{asc_cluster}
\end{figure}

\begin{figure}[tb]
\centering
\includegraphics[width=0.49\textwidth]{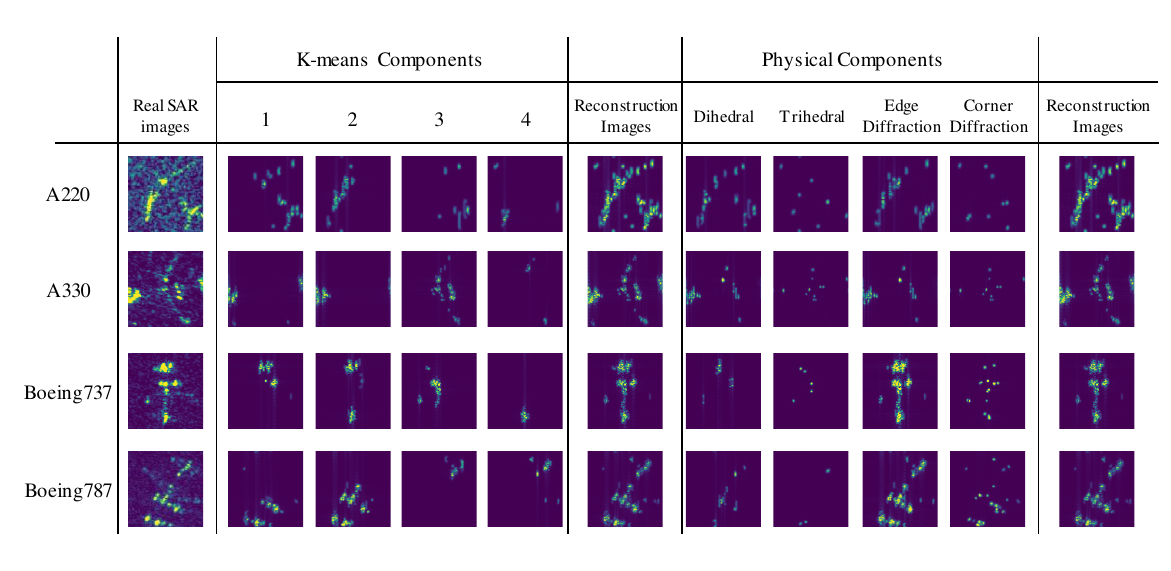}
\caption{ASCCs of different SAR aircraft targets and their corresponding reconstructed SAR images under two clustering methods. When clustered based on physical meaning, the different ASCCs exhibit distinct geometric types, like dihedral.}
\label{asc_cluster_air}
\end{figure}

\subsection{Validation of Method's Interpretability}
\label{experiments_validation}

This subsection aims to validate the interpretability and robustness of the proposed method. Section \ref{experiments_validation1} is designed to assess whether the ASCCs exhibit intra-class compactness and inter-class separability, which is a critical prerequisite for confirming their effectiveness as a basis for recognition.
Section \ref{experiments_validation2} focuses on validating the transformation of uninterpretable deep features into a linear combination of ASCC features, demonstrating that the method can accurately decompose deep features into ASCCs' features with minimal error.

\subsubsection{Validation of ASCCs' Interpretability for recognition}
\label{experiments_validation1}
This part of the experiment aims to validate ASCCs' effectiveness as an interpretable reasoning basis for recognition.
First, the ASCCs of SAR targets under two different clustering methods are presented.
Next, from a quantitative perspective, commonly used metrics, structural similarity index measure(SSIM), mean squared error (MSE), and multi-scale SSIM (MS-SSIM), are employed to measure the intra-class compactness of the ASCCs. 
The experimental setup is as follows:

\textbf{Qualitative validation for ASCCs' interpretability for recognition.}
The qualitative validation for the effectiveness of ASCCs in recognition primarily includes two parts. 
Firstly, ASCCs are clustered in two ways: one using the K-means algorithm to obtain $K_{asc}$ ASCCs, and the other based on the constraints of Frequency Dependence and Length from Table \ref{tab:asc_components}.
Then, the $K_{asc}$ ASCCs are visualized using t-SNE for dimensionality reduction, showcasing the feature distribution of ASCCs for different types of SAR targets in 2D space. In this experiment, $K_{asc} = 6$.
For MSTAR, all SAR image data used in this experiment are samples with a depression angle of 17 degrees, taken from the training samples listed in Table \ref{ttnumMSTAR}.
For SAR-AIRcraft, the SAR images are sampled from  Table \ref{saracd}.

The results of SAR ASCCs generated using two different clustering methods are shown in Fig. \ref{asc_cluster} and Fig. \ref{asc_cluster_air}, separately under MSTAR and SAR-AIRcraft dataset. This figures display the ASCCs for different types of SAR targets, along with their corresponding reconstructed SAR images. 

From the presentation above, it is clear that the SAR ASCCs with different physical meanings have shown the interpretability for recognition. 

\textbf{Quantitative validation for ASCCs' interpretability for recognition.}
The quantitative validation of the effectiveness of ASCCs from recognition is as following:
Commonly used metrics such as SSIM, MSE, and MS-SSIM are employed to evaluate the intra-class compactness of ASCCs. To manage the large dataset and computational demands, 100 samples and their corresponding ASCCs are randomly selected for each SAR target type in Table \ref{ttnumMSTAR}. Pairwise SSIM, MSE, and MS-SSIM values within the same class are calculated to assess the intra-class compactness.

\begin{figure}[htb]
\begin{center}
\subfigure[Intra-class MSE similarity of ASCCs.]{\includegraphics[width=0.75\linewidth]{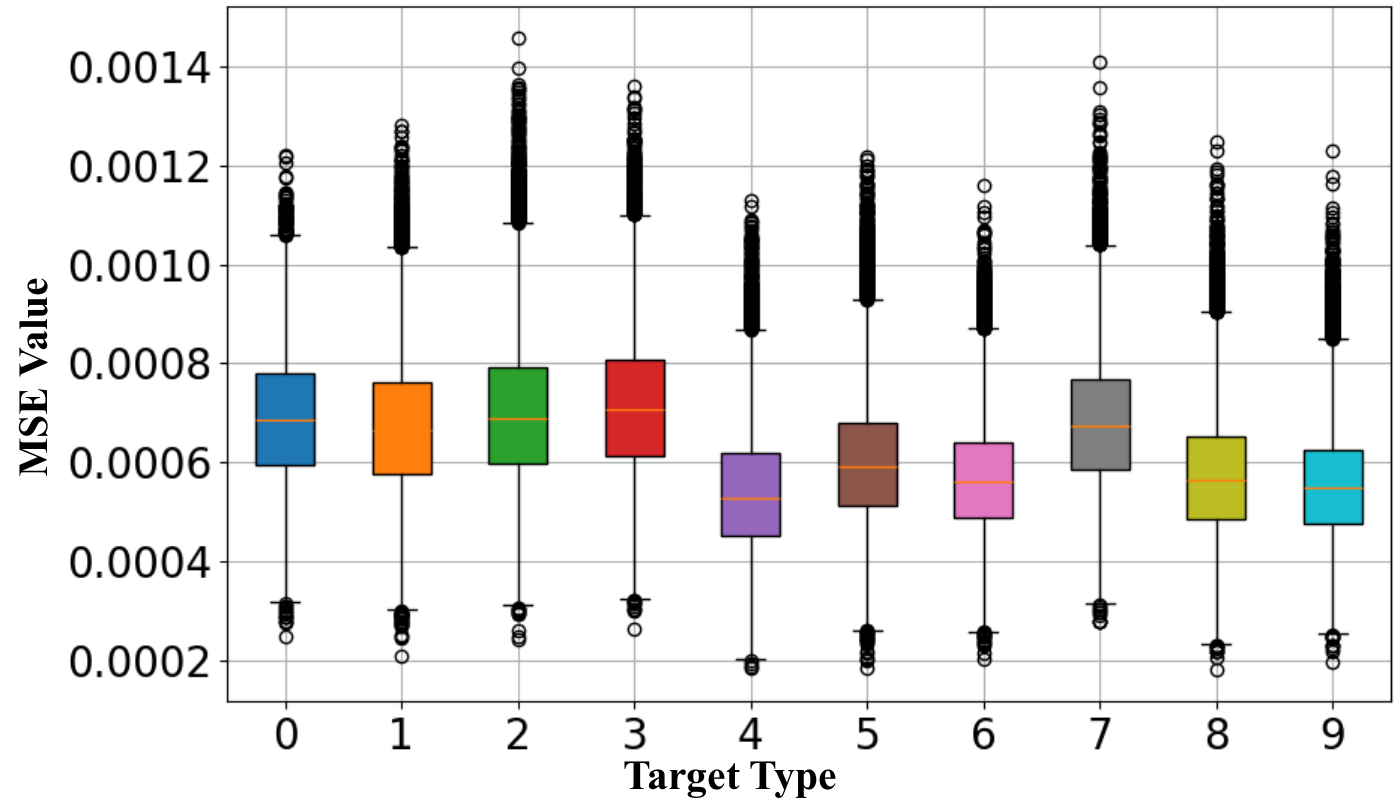}} \\
\subfigure[Intra-class SSIM similarity of ASCCs.]{\includegraphics[width=0.75\linewidth]{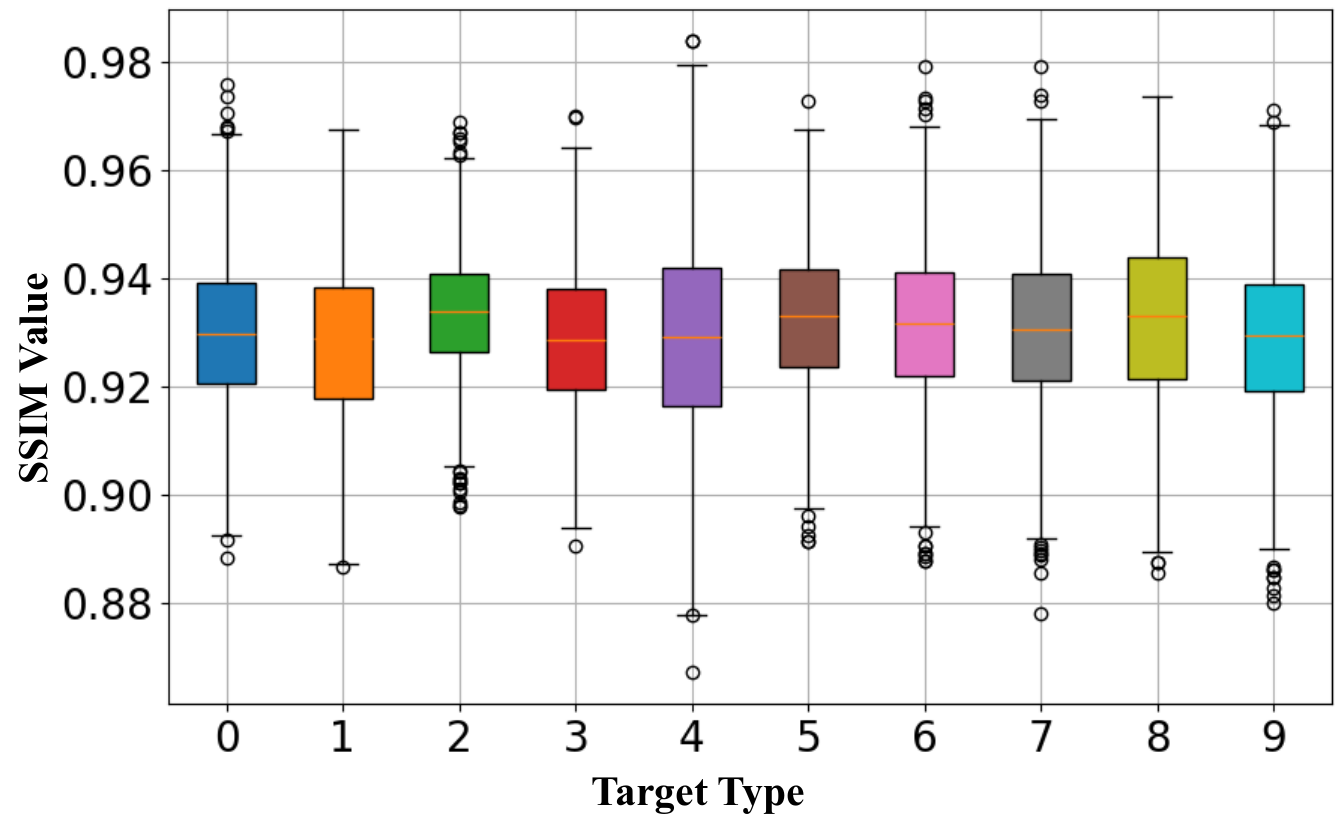}} \\
\subfigure[Intra-class similarity of ASCCs. The bubble size corresponds to MS-SSIM values.]{\includegraphics[width=0.75\linewidth]{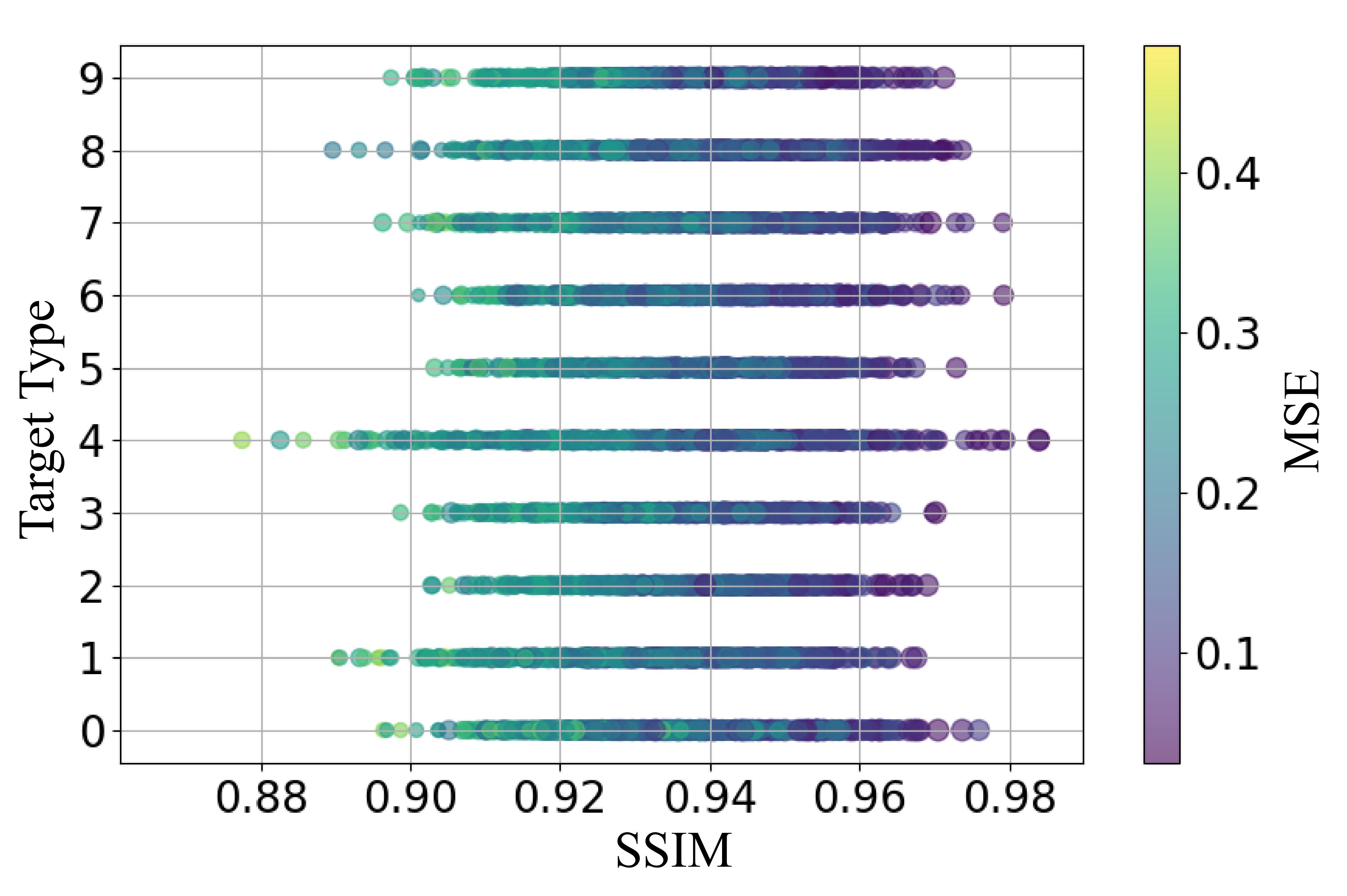}} 
\end{center}
\caption{Intra-class MSE, SSIM, and MSSSIM similarities of ASCCs to verify whether the ASCCs exhibit intra-class compactness.}
\label{ASC_visual_validation_inner}
\end{figure}

The intra-class compactness results are shown in Fig. \ref{ASC_visual_validation_inner}. Fig. \ref{ASC_visual_validation_inner}(a) presents the box plot of MSE values for ASCCs within the same class, computed for $K_{asc} \times K_{asc}$ pairs of SAR images and their corresponding $2 \times K_{asc}$ ASCCs. Fig. \ref{ASC_visual_validation_inner}(b) displays the box plot of SSIM values for ASCCs within the same class. Fig. \ref{ASC_visual_validation_inner}(c) presents a bubble chart where the x-axis represents SSIM, the y-axis indicates the category, color denotes MSE, and bubble size corresponds to MS-SSIM values.

From Fig. \ref{ASC_visual_validation_inner}, the MSE values for ASCCs within the same SAR images are generally below 0.0008, indicating minimal variation. The SSIM distribution shows that structural similarity is typically above 0.92. Fig. \ref{ASC_visual_validation_inner}(c) further demonstrates a clear relationship, where high SSIM values correspond to low MSE values.

Overall, the results demonstrate that ASCCs exhibit high intra-class compactness and inter-class separability, confirming their reliability as a robust reasoning basis for recognition.

\subsubsection{Validation of Method's Interpretability}
\label{experiments_validation2}
This subsection aims to validate the interpretability of the proposed method through two main experiments: (1) verifying the accuracy of decomposing deep features into ASCC features with minimal error, and (2) demonstrating distinct recognition logic when using $K_{asc}$ ASCCs to recognize different SAR target types. The experimental setup and results are as follows:

\textbf{Qualitative Validation of Decomposition Errors.}
The proposed method decomposes deep features into ASCC features in two stages. The first stage transforms deep features into all ASCCs, where decomposition errors play a critical role. The second stage further decomposes all ASCCs into individual ASCCs. To evaluate this, decomposition errors for both stages are separately recorded using SAR image data with a 15-degree depression angle, taken from the testing samples listed in Table \ref{ttnumMSTAR}. The decomposition error is measured using the squared Frobenius norm:

\begin{equation}\label{squared Frobenius norm}
\| X - UWV \|_F^2 = \sum_{i=1}^{r} \sum_{j=1}^{r} \left( X_{ij} - (UWV)_{ij} \right)^2
\end{equation}
where \( X \) is the original matrix, and \( U \), \( W \), and \( V \) are the decomposed factor matrices.

\begin{figure}[tb]
\begin{center}
\subfigure[Error distributions of different SAR target types in decomposing deep features into ASCCs at the first layer.]{\includegraphics[width=0.75\linewidth]{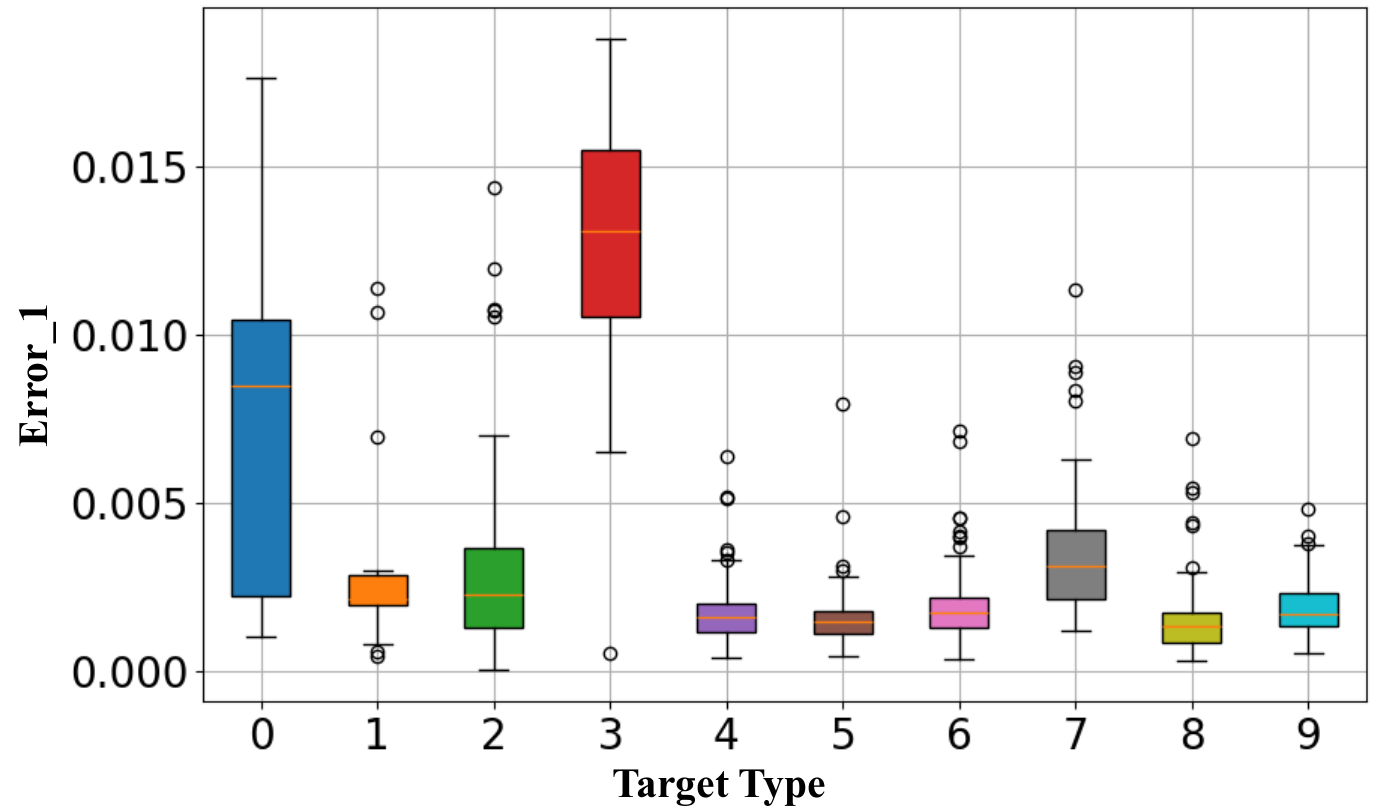}} 
\subfigure[Error distributions of different SAR target types in decomposing deep features into ASCCs from the 2nd to the k-th layer.]{\includegraphics[width=0.75\linewidth]{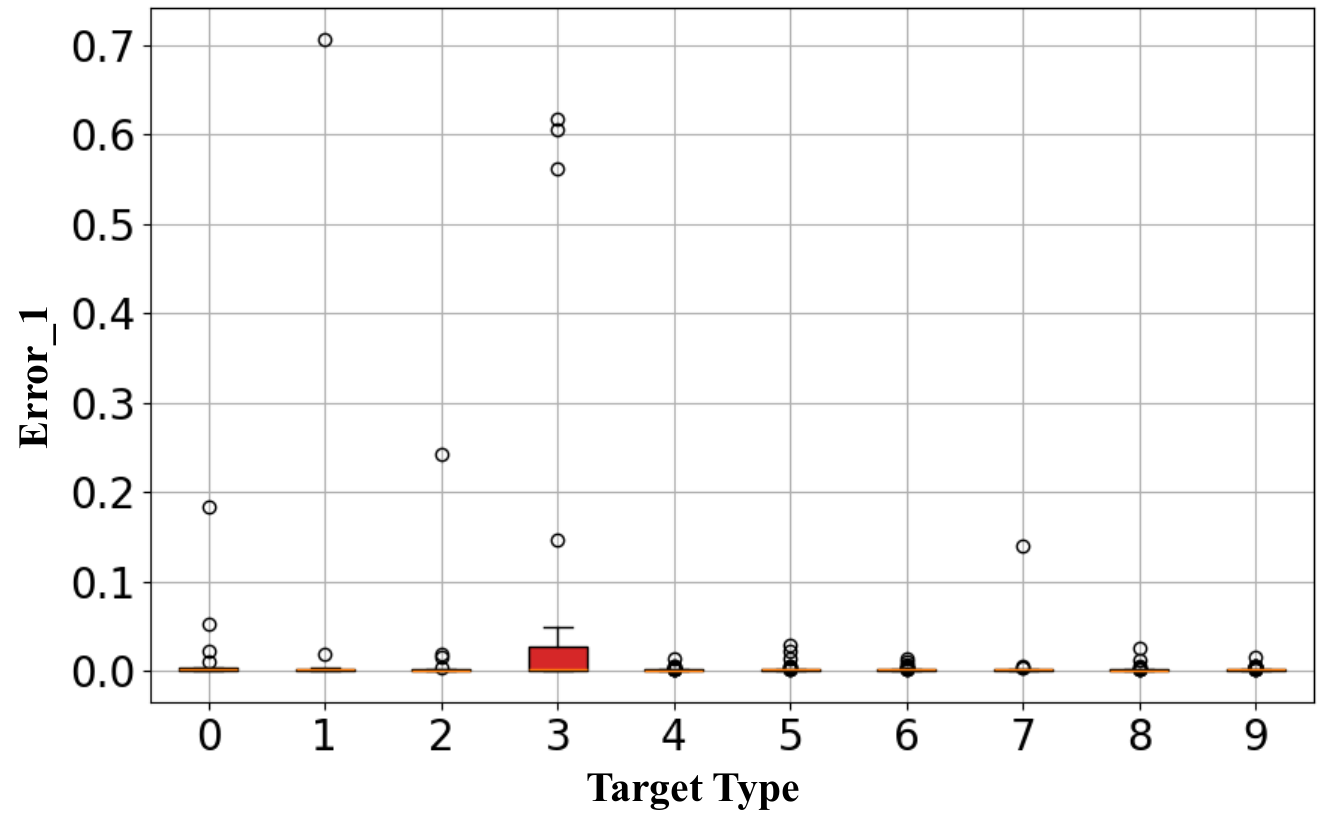}} \\
\subfigure[Error comparison of different SAR target types in decomposing deep features into ASCCs between the these two decompositions.]{\includegraphics[width=0.75\linewidth]{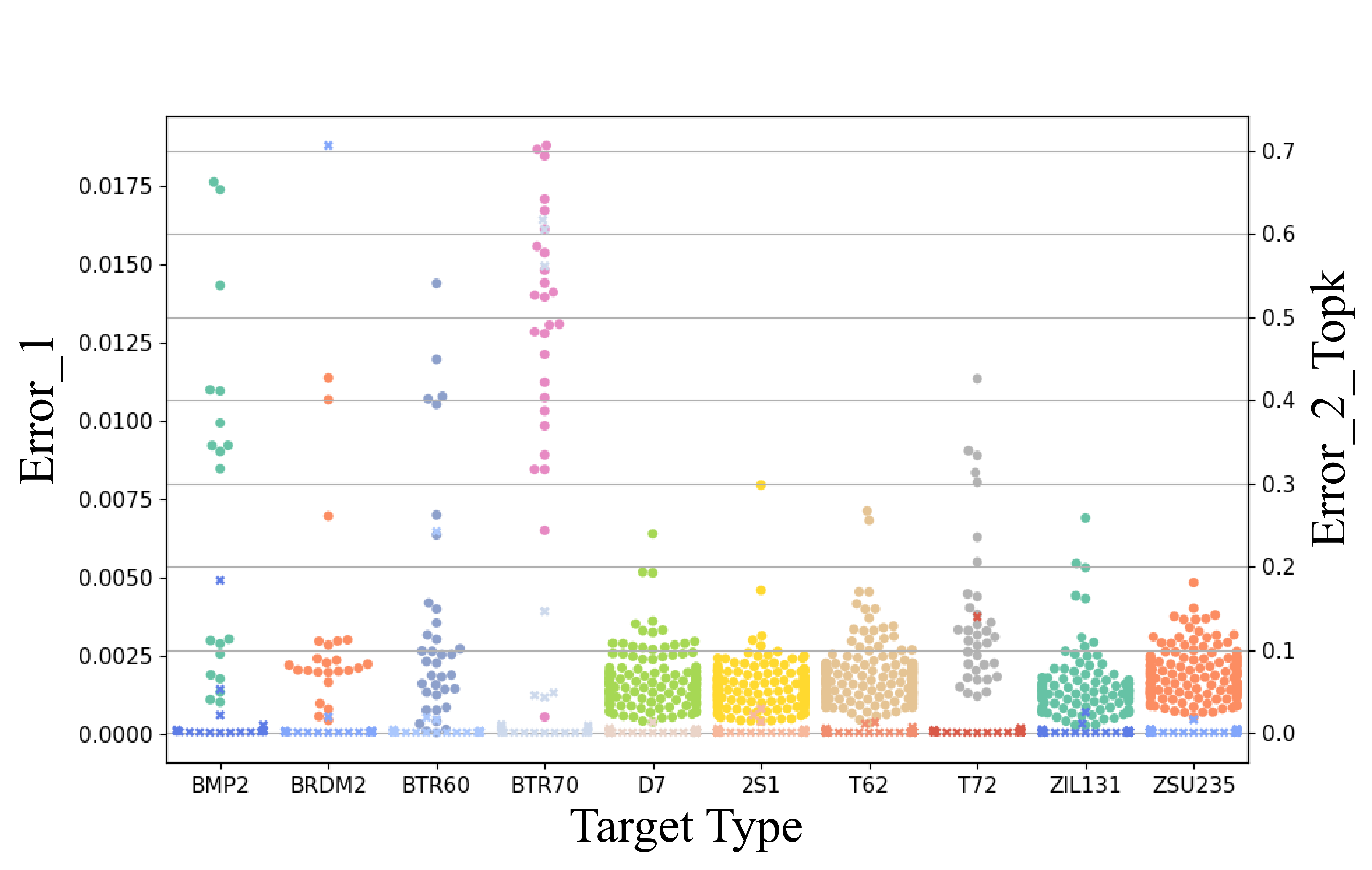}}
\end{center}
\caption{Proposed method's error in decomposing deep features into ASCCs. Since the proposed method involves decomposition at the first layer and from the 2nd to the k-th layer, the errors for these two decompositions are presented and compared separately.}
\label{ASC_dec_error_logic}
\end{figure}

The results, shown in Fig. \ref{ASC_dec_error_logic}, detail the error distributions for both stages. Subfigure (a) illustrates the error for different SAR target types in the first stage, while subfigure (b) presents the errors in the second stage. Subfigure (c) compares the errors across both stages, with the x-axis representing SAR target types, the left y-axis showing first-stage errors (solid circles), and the right y-axis indicating second-stage errors (crosses).

From Fig. \ref{ASC_dec_error_logic}(a), the first-stage errors remain consistently below 0.015. Although variance is higher for BMP2 and BTR70, the overall errors are low, with BMP2 showing a significant proportion of low errors (Fig. \ref{ASC_dec_error_logic}(c)). For the second stage, Fig. \ref{ASC_dec_error_logic}(b) and (c) demonstrate uniformly low errors across all target types. These findings indicate that the proposed method achieves low errors in the first-stage decomposition for most SAR target types, with negligible errors in the second stage. 

Thus, the method can effectively and accurately decompose deep features into ASCC features with minimal error.

\begin{figure}[tb]
\centering
\includegraphics[width=0.4\textwidth]{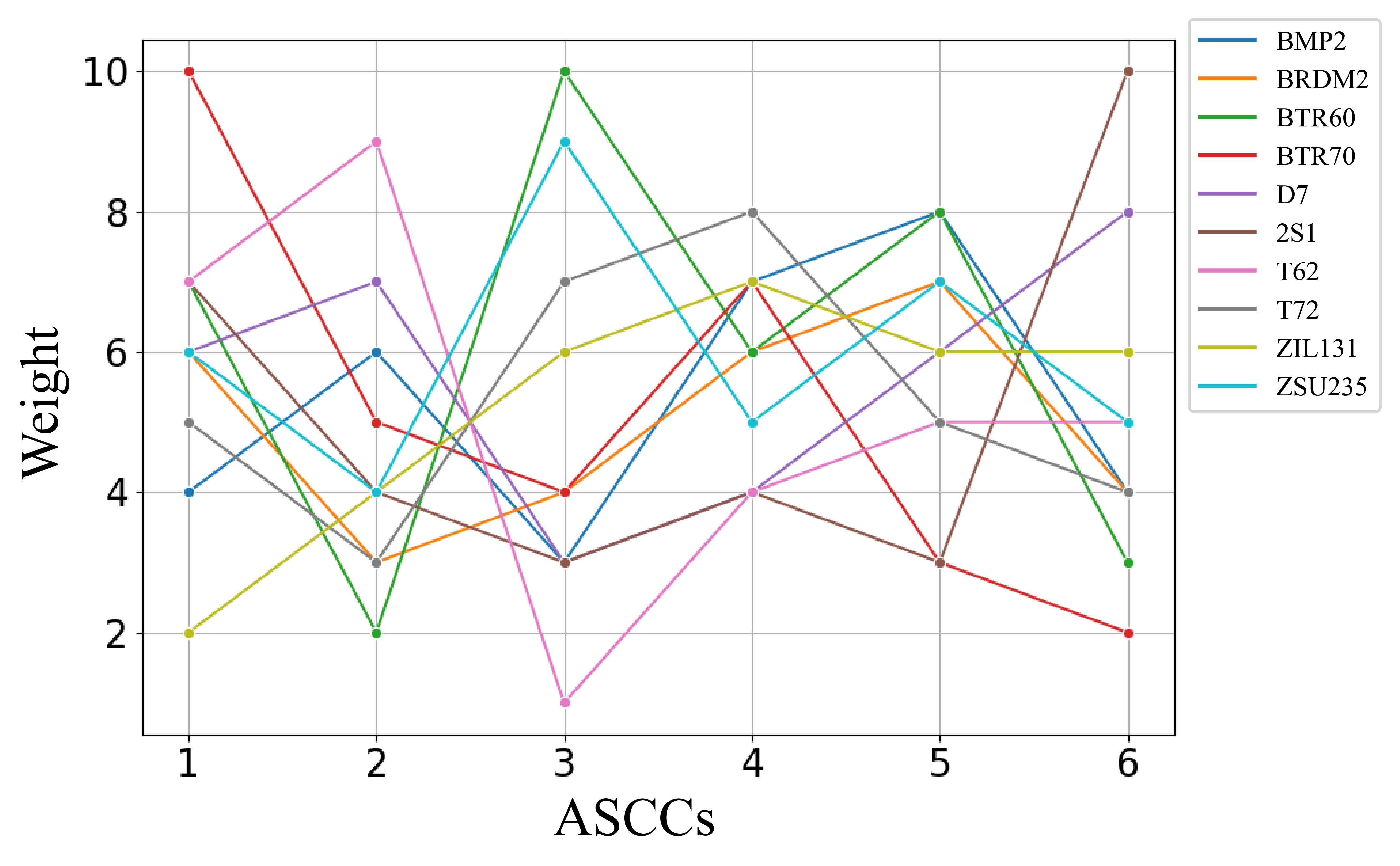}
\caption{Recognition logic of the proposed method for different SAR target types is illustrated, where the x-axis represents different ASCCs. The weights, normalized between 1 and 10, are shown on the y-axis, with different colored lines representing different SAR target types.}
\label{recognition_logic_line}
\end{figure}

\textbf{Verification of Recognition Logic Using ASCC Weights.}
To further validate the method's interpretability, the recognition logic is visualized by analyzing the weights assigned to ASCCs. During recognition, the proposed method inputs $K_{asc}+1$ parameters $W_i$ into the classifier to identify SAR target types. These parameters are then decomposed into $K_{asc}$ ASCCs. This experiment calculates the weights assigned to the ASCCs, normalizes them to the range \([1, 10]\), and visualizes the weight distributions for different SAR target types (Fig. \ref{recognition_logic_line}).

As shown in Fig. \ref{recognition_logic_line}, the method demonstrates significant variation in recognition logic for different SAR target types when $K_{asc}=6$, reflecting distinct weight distributions across the ASCCs.

\renewcommand{\arraystretch}{1.2}
\begin{table*}
    \centering
    \footnotesize
    \caption{Recognition Performances of 10 Classes under Different Training Data in MSTAR}
    \label{mstar10-r}
    \small
        \begin{tabular}{c|cccccccccccccc}
            \toprule \toprule
            \multirow{2}{*}{Class} & \multicolumn{13}{c}{Training Number in Each Class}                                                                                                         \\\cline{2-14}
                                   & 5                                                  & 10    & 20     & 25    & 30     & 35    & 40     & 50    & 60     & 70     & 80     & 100    & 220    \\ \midrule
            BMP2                   & 63.59                                              & 75.90 & 97.85  & 92.31 & 96.89  & 96.52 & 93.85  & 97.95 & 98.46  & 98.48  & 100.00 & 99.49  & 100.00 \\
            BRDM2                  & 86.50                                              & 91.61 & 97.12  & 95.26 & 97.15  & 97.10 & 97.81  & 95.26 & 99.27  & 99.27  & 100.00 & 99.27  & 100.00 \\
            BTR60                  & 77.44                                              & 79.49 & 93.33  & 90.77 & 94.74  & 94.61 & 93.33  & 97.95 & 96.41  & 97.96  & 98.97  & 99.49  & 99.49  \\
            BTR70                  & 60.20                                              & 85.20 & 92.19  & 90.82 & 93.53  & 95.02 & 98.98  & 98.98 & 97.45  & 98.47  & 99.49  & 100.00 & 100.00 \\
            D7                     & 72.99                                              & 97.81 & 100.00 & 98.18 & 99.61  & 98.91 & 94.89  & 99.27 & 99.64  & 99.63  & 100.00 & 100.00 & 100.00 \\
            2S1                    & 87.59                                              & 90.88 & 100.00 & 97.45 & 100.00 & 98.42 & 95.26  & 99.27 & 100.00 & 100.00 & 98.54  & 99.27  & 100.00 \\
            T62                    & 90.84                                              & 97.07 & 97.82  & 99.27 & 97.75  & 99.26 & 99.27  & 99.63 & 99.27  & 99.27  & 100.00 & 100.00 & 100.00 \\
            T72                    & 66.33                                              & 98.47 & 92.27  & 96.94 & 95.41  & 96.02 & 100.00 & 98.98 & 97.96  & 96.98  & 100.00 & 100.00 & 100.00 \\
            ZIL131                 & 93.43                                              & 97.81 & 96.47  & 99.64 & 94.14  & 99.61 & 99.64  & 98.54 & 98.91  & 99.64  & 100.00 & 100.00 & 100.00 \\
            ZSU234                 & 92.34                                              & 99.64 & 94.48  & 99.63 & 97.16  & 96.14 & 100.00 & 99.64 & 98.91  & 99.28  & 99.64  & 100.00 & 100.00 \\ \midrule
            Average                & 80.70                                              & 92.25 & 96.37  & 96.45 & 96.78  & 97.32 & 97.40  & 98.56 & 98.76  & 99.01  & 99.67  & 99.75  & 99.96  \\ \bottomrule \bottomrule
        \end{tabular}
\end{table*}

\renewcommand{\arraystretch}{1.2}
\begin{table*}
    \centering
    \footnotesize
    \caption{Recognition Performances (\%) of 6 Classes under Different Training Data in OpenSARShip}
    \label{opensarship6-r}
    \small
        \begin{tabular}{c|ccccccccccc}
            \toprule \toprule
            \multirow{2}{*}{Class} & \multicolumn{11}{c}{Training Number in Each Class}                                                                                 \\ \cline{2-12}
                                   & 10                                                 & 20    & 30    & 40    & 50    & 60    & 70    & 80    & 90    & 100   & 200   \\ \midrule
            Bulk Carrier           & 99.79                                              & 74.74 & 99.79 & 74.74 & 74.74 & 99.58 & 99.58 & 99.58 & 94.67 & 99.58 & 99.79 \\
            Container Ship         & 99.38                                              & 84.59 & 73.98 & 84.59 & 84.59 & 99.38 & 84.59 & 99.36 & 99.45 & 99.38 & 99.38 \\
            Tanker                 & 0.00                                               & 67.80 & 67.80 & 67.80 & 67.80 & 33.90 & 67.80 & 40.82 & 67.80 & 67.80 & 67.80 \\
            Cargo                  & 71.27                                              & 71.81 & 71.81 & 78.46 & 76.92 & 71.27 & 71.81 & 73.07 & 78.46 & 78.46 & 99.82 \\
            Fishing                & 31.40                                              & 31.40 & 31.40 & 31.40 & 50.67 & 31.40 & 31.40 & 25.17 & 31.40 & 31.40 & 31.40 \\
            General Cargo          & 0.00                                               & 0.00  & 0.00  & 0.00  & 33.47 & 0.00  & 0.00  & 26.67 & 0.00  & 0.00  & 0.00  \\ \midrule
            Average                & 69.07                                              & 69.23 & 70.56 & 70.72 & 72.57 & 73.86 & 73.98 & 75.47 & 79.58 & 80.31 & 85.14 \\ \bottomrule \bottomrule
        \end{tabular}
\end{table*}

\renewcommand{\arraystretch}{1.2}
\begin{table}
    \centering
    \caption{Recognition Performances (\%) of 5 Classes under Different Training Data in FuSAR-Ship}
    \label{fusar-r}
    \resizebox{\linewidth}{!}{
        \begin{tabular}{c|ccccccc}
            \toprule \toprule
            \multirow{2}{*}{Class} & \multicolumn{7}{c}{Training Number in Each Class}                                                 \\ \cline{2-8}
                                   & 10                                                & 20    & 30    & 40    & 60    & 70    & 100   \\ \midrule
            BulkCarrier            & 99.58                                             & 66.11 & 31.31 & 51.67 & 43.33 & 63.33 & 60.56 \\
            CargoShip              & 99.83                                             & 70.00 & 87.50 & 81.52 & 83.33 & 75.95 & 80.51 \\
            Fishing                & 59.64                                             & 43.01 & 57.14 & 43.00 & 66.00 & 63.33 & 65.00 \\
            Other type of ship     & 55.56                                             & 91.50 & 98.80 & 95.30 & 93.62 & 98.33 & 95.30 \\
            Tanker                 & 22.87                                             & 0.00  & 25.42 & 96.72 & 0.00  & 40.30 & 20.53 \\ \midrule
            Average                & 61.00                                             & 66.45 & 67.89 & 72.06 & 72.28 & 78.00 & 78.21 \\ \bottomrule \bottomrule
        \end{tabular}
    }
\end{table}

\renewcommand{\arraystretch}{1.2}
\begin{table*}
    \centering
    \footnotesize
    \caption{Recognition Performances (\%) of 6 Classes under Different Training Data in SAR-AIRcraft}
    \label{saracd6-r}
    \small
        \begin{tabular}{c|cccccccccccc}
            \toprule \toprule
            \multirow{2}{*}{Class} & \multicolumn{12}{c}{Training Number in Each Class}                                                                                         \\\cline{2-13}
                                   & 10                                                 & 20    & 30    & 40    & 50    & 60    & 70    & 80    & 90    & 100   & 110   & 150   \\ \midrule
            A220                   & 40.00                                              & 65.42 & 76.67 & 99.72 & 76.11 & 75.83 & 67.33 & 81.67 & 77.92 & 98.33 & 64.00 & 87.22 \\
            A330                   & 40.00                                              & 52.33 & 49.29 & 55.67 & 59.00 & 48.33 & 70.56 & 82.08 & 99.17 & 65.67 & 99.17 & 87.22 \\
            A320/321                & 99.17                                              & 72.29 & 79.44 & 94.17 & 90.83 & 96.11 & 72.08 & 62.67 & 78.33 & 99.58 & 84.58 & 87.78 \\
            ARJ21                  & 66.67                                              & 91.67 & 58.33 & 80.00 & 83.33 & 99.72 & 91.67 & 75.00 & 88.89 & 99.17 & 99.72 & 88.89 \\
            Boeing737              & 88.33                                              & 70.42 & 85.00 & 64.67 & 71.67 & 85.56 & 89.17 & 99.58 & 66.33 & 87.08 & 93.33 & 93.89 \\
            Boeing787              & 79.02                                              & 75.00 & 99.74 & 76.56 & 99.40 & 76.56 & 99.26 & 99.40 & 99.70 & 73.45 & 99.74 & 99.70 \\ \midrule
            Average                & 63.53                                              & 70.34 & 70.31 & 74.83 & 78.56 & 79.89 & 80.16 & 81.70 & 81.74 & 83.26 & 87.23 & 90.67 \\ \bottomrule \bottomrule
        \end{tabular}
\end{table*}

\subsubsection{Ablation Experiments}

This ablation experiments aim to validate the intra-class compactness and inter-class separability of the ASCCs, which is crucial to establish their effectiveness as an interpretable reasoning basis for recognition.

First, the feature distribution of ASCCs for different types of SAR targets is visualized in a 2D space using t-SNE to verify their inter-class separability. Then, cosine similarity is used to measure the inter-class separability of ASCCs. Similarly, 100 samples and their corresponding ASCCs are randomly selected from each category in Table \ref{ttnumMSTAR}, and cosine similarity values are computed between ASCC pairs from different classes to validate inter-class separability.

\begin{figure}[htb]
\begin{center}
\subfigure[the distribution of ASCCs.]{\includegraphics[width=0.45\linewidth]{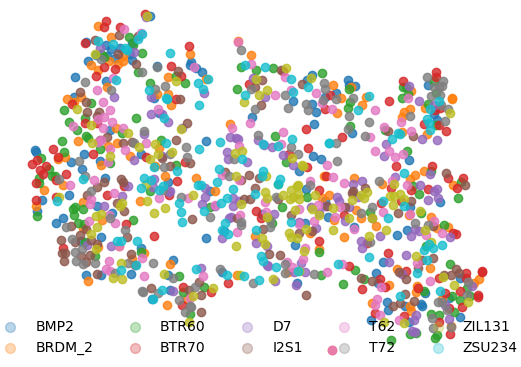}} 
\subfigure[the distribution of ASCCs' features.]{\includegraphics[width=0.45\linewidth]{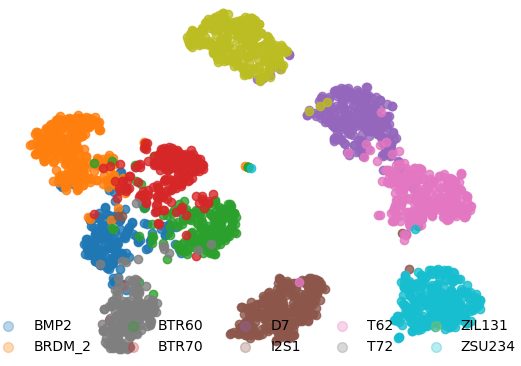}}
\end{center}
\caption{Distribution of ASCCs and their features in 2D space after dimensionality reduction using t-SNE. It is clear that the features of ASCCs exhibit stronger intra-class compactness and inter-class separability compared to the ASCCs. }
\label{asc_feadistribution}
\end{figure}

The feature distributions of ASCCs for various SAR target types are illustrated in Fig. \ref{asc_feadistribution}, where different colors represent different SAR target types, with specific correspondences labeled in the figure.

The experiments validate the effectiveness of using ASCCs as the recognition basis and the accuracy of the proposed method in decomposing deep features into ASCCs. Moreover, the method provides interpretable recognition logic, enabling not only accurate recognition but also a physically meaningful and transparent reasoning process.

\subsection{Recognition Results under Four Benchmark Datasets}

\subsubsection{Recognition Results under MSTAR}\label{sec: Recognition Results under MSTAR}

We evaluate the recognition performance of the proposed method on the MSTAR dataset, which includes ten different target classes. Training samples are collected at a $\text{17}^\circ$ depression angle, while testing samples are collected at a $\text{15}^\circ$ depression angle. The distribution of training and testing images is shown in Table~\ref{ttnumMSTAR}, with the numbers representing the original SAR image counts in the dataset.

Table~\ref{mstar10-r} summarizes the SOC recognition performance of the proposed method. The first row specifies the number of training samples per target class, while the first column lists the target classes and their corresponding series. Recognition ratios for each class and the overall average are based on 20 experimental runs.

The results reveal that with at least 20 training samples per class, recognition rates exceed 96.00\%. When the number of training samples per class increases to over 70, recognition rates improve further, surpassing 99.00\%. However, with only 5-10 training samples per class, recognition rates show greater variability, underscoring the challenges posed by limited sample sizes. Notably, under the demanding 5-shot scenario (50 training images before augmentation), the method achieves a strong recognition accuracy of 79.26\% for ten-class recognition.

These results demonstrate that the proposed method delivers robust recognition performance across varying sample sizes, from as few as 5 to as many as 200 training samples per target class in the MSTAR dataset.

\subsubsection{Recognition Results under OpenSARship}\label{sec: Recognition Results of 3 and 6 classes under OpenSARship}

The OpenSARShip dataset includes several ship classes representing a significant portion of the global shipping market \cite{intro_aug_consit2}. Following previous studies \cite{open2}, the 6-class experiment includes bulk carriers, container ships, tankers, cargo ships, fishing vessels, and general cargo ships. The results are presented in Tables~\ref{opensarship6-r} by varying the number of training samples per class from 10 to 200.

The results demonstrate the effectiveness of the proposed method in recognizing SAR ship images, showing strong robustness under varying training sample sizes. In the experiment, when the number of training samples per class decreases from 40 to 10, the recognition rate drops by only 1.65\%, from 70.72\% to 69.07\%. Similarly, reducing training samples from 80 to 40 results in a slightly larger decline of 4.75\%, from 75.47\% to 70.72\%. Further reduction from 200 to 100 samples per class leads to a 4.83\% decrease, from 85.14\% to 80.31\%, underscoring the robustness of the method across different numbers of training samples.

These findings highlight the robustness of the proposed method in scenarios with limited training ship data, as it maintains high recognition performance with only marginal decreases under reduced sample conditions. 

\subsubsection{Recognition Results under FUSAR-Ship}\label{sec: Recognition Results under FUSAR-Ship}

The recognition task on the FUSAR-Ship dataset poses greater complexity compared to the OpenSARShip dataset. The FUSAR-Ship dataset includes five major ship classes common in the global shipping market, along with an additional class labeled "other types of ships." This class encompasses a wide variety of ship types beyond the primary five, introducing increased overlap and demanding greater robustness and efficiency from the proposed method. The original training and testing set composition for the FUSAR-Ship dataset is detailed in Table~\ref{fusarset}.

Table~\ref{fusar-r} summarizes the recognition performance of the proposed method on the FUSAR-Ship dataset across five classes, with the number of training samples per class ranging from 100 to 10. The method consistently demonstrates strong performance. When training samples decrease from 100 to 40, the recognition rate drops by 6.15\%, from 78.21\% to 72.06\%. As the number of training samples reduces further, from 40 to 10, the recognition rate declines by 11.06\%, from 72.06\% to 61.00\%, reflecting the added complexity of the dataset.

\subsubsection{Recognition Results under SAR-AIRcraft}\label{sec: Recognition Results under AIRcraft}
The recognition tasks for SAR-AIRcraft dataset are conducted under six classes as detailed in Table~\ref{saracd}. The number of training samples are randomly selected from per class, then the rest samples are treated as testing samples. With an average for 20 experiments, the recognition performances under different training samples are shown in Table~\ref{saracd6-r}.

From the results, we can see that our proposed method can also be well adapted to the more complicated SAR aircraft targets recognition. In this experiment, when there are only 10 samples in per target class, the recognition accuracy can reach 63.53\%. With the training number increasing, the recognition performance is improved obviously. From 10 to 20 samples per class can lead to a 6.81\% increase, from 63.53\% to 70.34\%. And when the number of training samples in per class is above 70, the method can achieve a performance of over 80\%. When the training number is 150 in per class, the model can obtain a strong performance of 90.67\%. These results demonstrate that our methods also have a satisfactory applicability for SAR aircraft targets recognition.

The method's performance across the MSTAR, OpenSARShip, FUSAR-Ship and SAR-Aircraft datasets highlights its effectiveness and robustness under limited training sample conditions. Its ability to handle diverse imaging scenes, target types (including vehicles, ships and aircraft), and complex imaging conditions demonstrates the versatility and adaptability of the approach, making it well-suited for a wide range of practical applications.

\subsection{Validation of Method's Soundness}
\label{experiments_recognition}

This subsection validates the recognition performance of the proposed method under various conditions: (1) Different backbones to ensure the method's interpretable recognition logic across architectures and size; (2) Varying numbers of ASCCs to assess effectiveness, given the time-consuming nature of ASCCs' acquisition.

\begin{figure}[tb]
\includegraphics[width=0.9\linewidth]{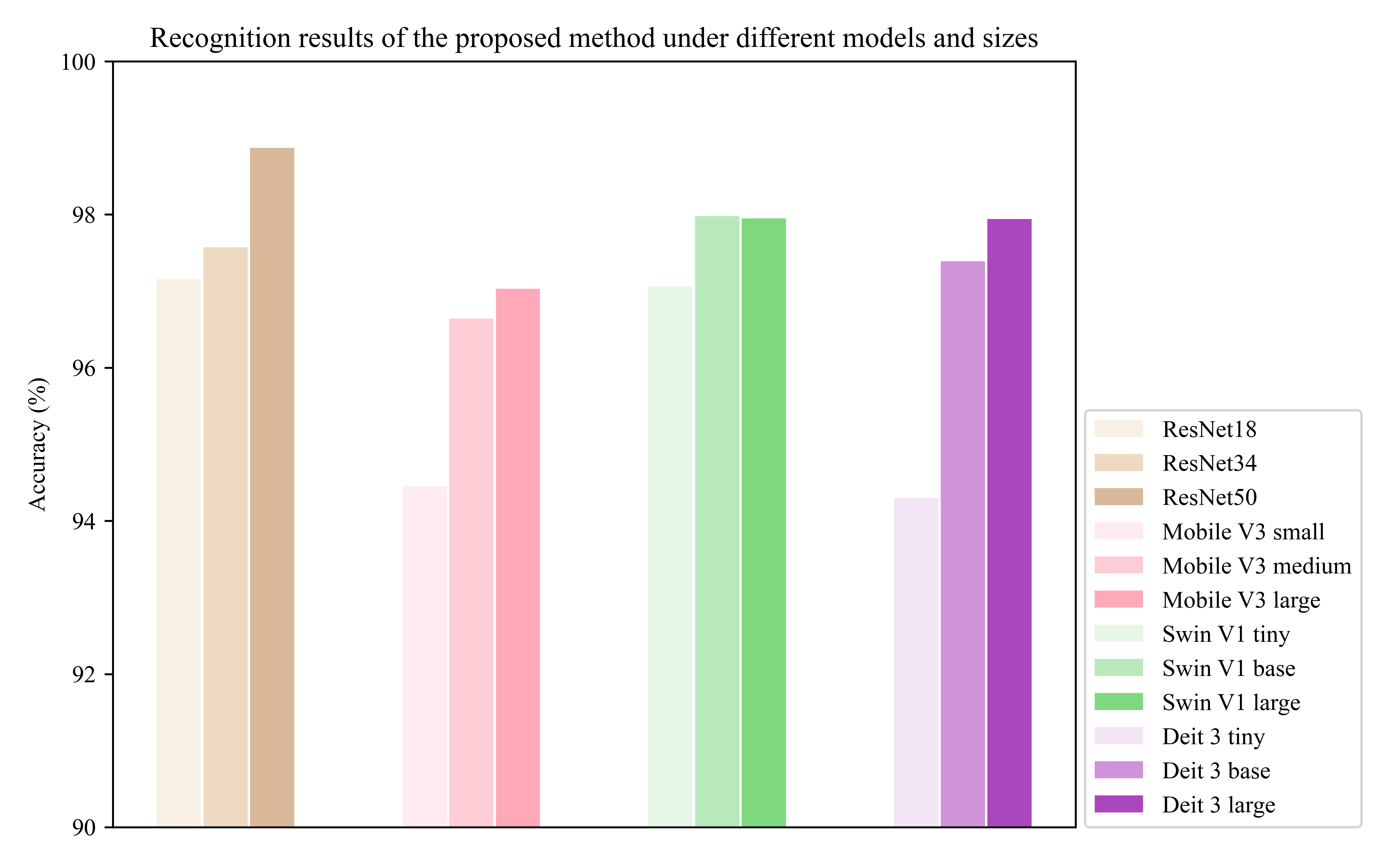}
\caption{Recognition results of the proposed method across various common deep learning architectures of different scales.}
\label{experiment_recogniton_arch}
\end{figure}

\begin{figure}[tb]
\includegraphics[width=0.9\linewidth]{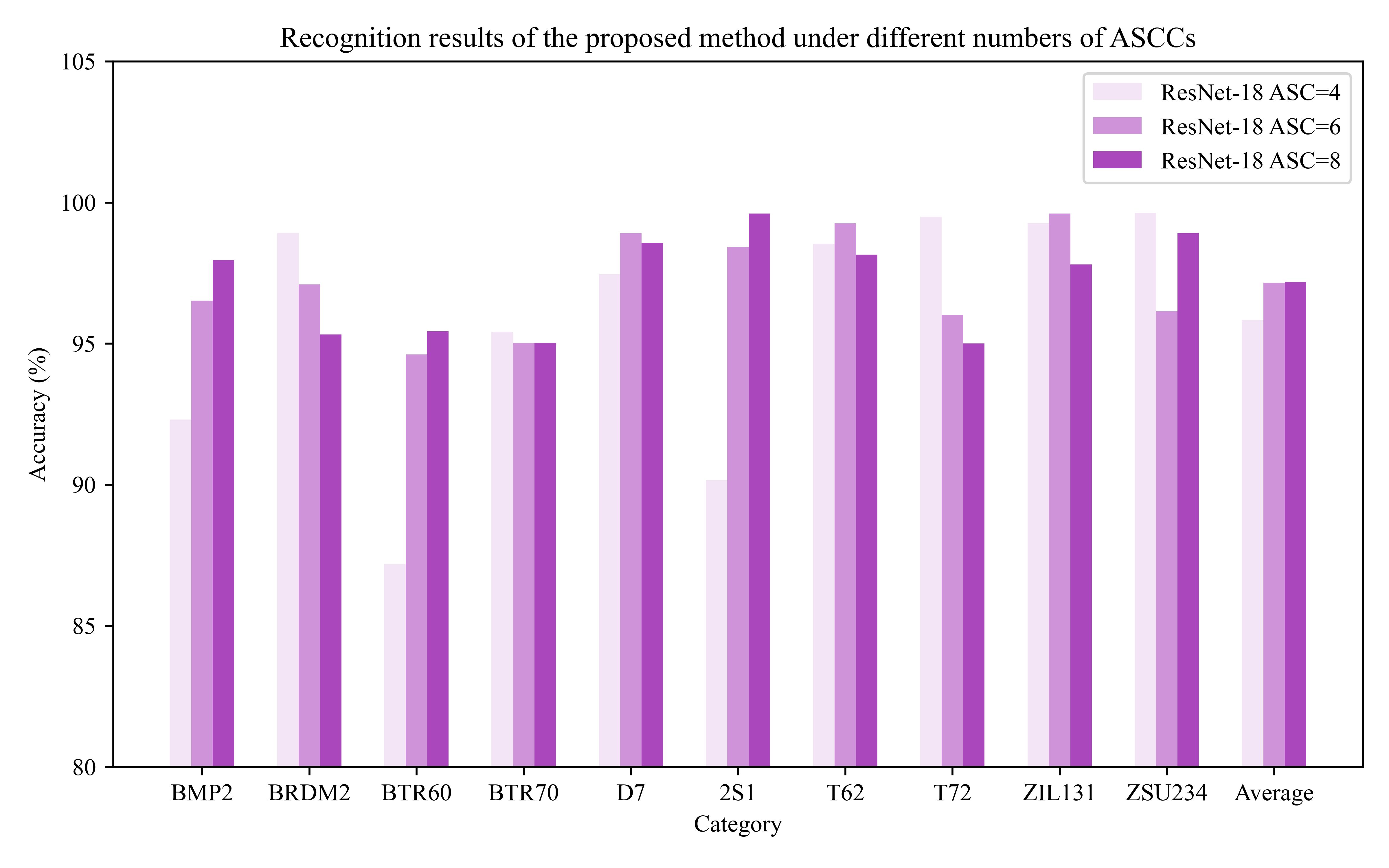}
\caption{Recognition results of the proposed method with different numbers of ASCCs.}
\label{recognition_asccs_number}
\end{figure}

Firstly, we validate the generalization ability of our method across various backbones by conducting recognition performance experiments with both CNN-based and transformer-based architectures. The recognition results are shown in Fig.~\ref{experiment_recogniton_arch}. It can be observed that under the training conditions of $K_{asc}=6$, our method achieves accurate recognition across different model size and architectures. 

In practical applications, acquiring ASCCs for SAR targets can be time-consuming. While some deep learning methods aim to accelerate this process, real-time performance remains a challenge. To evaluate the recognition performance of our method with varying numbers of ASCCs, we conducted experiments using ASCC quantities of 4, 6, and 8, with ResNet-18 and Swin-Tiny for performance validation. The experimental settings were consistent with previous experiments.

The recognition results with varying numbers of ASCCs are illustrated in Fig.~\ref{recognition_asccs_number}. As can be observed, when the number of ASCCs is 6 or 8, the average recognition performance based on ResNet-18 surpasses 95.00\%, which is superior to the performance when the number of ASCCs is 4. This suggests that while increasing the number of ASCCs can enhance performance, the improvement is marginal beyond a certain point. Therefore, around 6 ASCCs represents an optimal choice for balancing recognition accuracy with the time and resources required for ASCC acquisition.

These experiments demonstrate that our method achieves accurate recognition using different deep learning architectures, including lightweight models that are well-suited for resource-constrained environments. Moreover, even with a few number of ASCCs, the method maintains accurate recognition, ensuring both practical applicability and real-time performance.

\subsection{Comparison of Recognition performance}
\label{experiments_comparison}

\renewcommand{\arraystretch}{1.2}
\begin{table*}
\centering
\footnotesize
\caption{{Comparison with Effective Deep Learning Methods and Specific Methods for SAR ATR Under Constant Training Samples of OpenSARShip. (The Training Samples of Methods Under 3 Classes and 6 Classes Are Evaluated Under 338 and 200 Samples Each Class.)}}
\label{3&6COMPARISONOpenSARShip}
\setlength\tabcolsep{1.5pt}
\begin{tabular}{c|c|cccccccc}
\toprule \toprule 
\multicolumn{2}{c|}{\multirow{2}{*}{Model}} & \multicolumn{4}{c}{3 classes}                                                                                                                                                                                                                                                 & \multicolumn{4}{c}{6 classes}                          \\ \cline{3-10} 
\multicolumn{2}{c|}{}    & Recall (\%)                                                       & Precision (\%)                                                    & F1 (\%)                                                           & Acc (\%)                                                          & Recall (\%) & Precision (\%) & F1 (\%)    & Acc (\%)   \\ \midrule
\multirow{15}{*}{\begin{tabular}[c]{@{}c@{}}Effective \\ deep learning \\ methods \end{tabular}} & 
VGG-11 \cite{2compared3} & 73.21±0.96 & 68.64±1.49 & 70.85±1.07 & 73.42±0.75 & 51.38±0.82 & 43.67±1.23 & 47.21±1.25 & 49.41±0.99 \\
& GooLeNet \cite{2compared5} & 69.73±2.70 & 68.80±1.81 & 69.21±1.19 & 73.80±1.32 & 54.47±0.95 & 44.96±1.76 & 49.25±0.70 & 49.76±1.56 \\
& ResNet-18 \cite{2compared6} & 73.76±1.61 & 69.40±1.92 & 71.49±1.04 & 74.64±0.68 & 50.19±0.47 & 42.85±1.20 & 46.23±0.35 & 45.91±0.43 \\
& ResNet-50 \cite{2compared6} & 71.67±1.71 & 66.79±1.27 & 69.13±1.04 & 72.82±0.75 & 50.27±1.21 & 43.32±1.32 & 46.54±2.50 & 49.80±1.70 \\
& DenseNet-161 \cite{2compared10} & 72.54±3.39 & 67.77±1.46 & 70.02±1.51 & 73.39±0.79 & 54.98±0.82 & 47.57±0.82 & 51.01±1.63 & 54.27±3.41 \\
& MobileNet-v3-Large \cite{2compared12}     & 65.12±2.53 & 60.75±1.72 & 62.84±1.73 & 66.13±0.92 & 49.95±0.58 & 42.14±0.62 & 45.71±0.65 & 46.60±2.61 \\
& MobileNet-v3-Small \cite{2compared12} & 67.23±1.59 & 61.85±1.69 & 64.42±1.41 & 66.71±0.87 & 48.28±0.75 & 40.75±0.73 & 44.20±0.57 & 44.41±1.10 \\
& SqueezeNet \cite{2compared13}    & 71.47±1.31 & 66.73±1.70 & 69.01±1.28 & 72.15±1.25 & 53.24±0.75 & 45.55±0.79 & 49.10±0.85 & 53.12±1.12 \\
& Inception-v4 \cite{2compared14}           & 69.26±3.16 & 67.43±2.39 & 68.28±1.97 & 72.44±0.70 & 54.92±0.69 & 46.46±0.49 & 50.34±1.31 & 54.55±3.52 \\
& Xception \cite{2compared15} & 71.56±3.00 & 68.60±1.67 & 70.00±1.29 & 73.74±0.86 & 52.21±0.94 & 44.03±1.15 & 47.77±1.11 & 49.56±1.47 \\ \midrule
\multirow{8}{*}{\begin{tabular}[c]{@{}c@{}}Methods for \\ Limited-data \\ SAR ATR \end{tabular}}  &  Wang et al. \cite{p4}            & 57.72±1.37 & 58.72±4.76 & 58.12±2.67 & 69.27±0.27 & 50.53±1.85 & 41.77±1.34 & 45.73±2.48 & 48.43±3.71 \\
& Hou et al. \cite{FUSAR}             & 69.33±2.00 & 69.44±2.42 & 66.76±1.64 & 67.41±1.13 & 48.76±0.79 & 41.22±0.74 & 44.67±1.21 & 47.44±2.01 \\
& Huang et al. \cite{2compared18}           & 74.74±1.60 & 69.56±2.38 & 72.04±1.60 & 74.98±1.46 & 54.09±0.81 & 47.58±1.66 & 50.63±1.79 & 54.78±2.08 \\
& Zhang et al. \cite{open3}           & 77.87±1.14 & 73.42±1.06 & 75.05±1.10 & 78.15±0.57 & 54.20±1.09 & 46.66±1.07 & 50.15±1.24 & 53.77±3.63 \\
& Zeng et al. \cite{OpenSARShip}            & 74.99±1.55 & 74.05±1.75 & 74.52±1,02 & 77.41±1.74 & 55.66±1.23 & 47.16±1.70 & 50.96±1.18 & 55.26±2.36 \\
& Xiong et al. \cite{2compared21}          & 73.87±1.16 & 71.50±3.00 & 72.67±2.04 & 75.44±2.68 & 53.57±0.33 & 45.74±0.82 & 49.35±0.69 & 54.93±2.61 \\
& SF-LPN-DPFF \cite{2compared22}           & 78.83±1.32 & 76.45±1.16 & 77.62±1.23 & 79.25±0.83 & 54.49±0.70 & 48.61±1.32 & 51.38±1.26 & 56.66±1.54 \\ \midrule
\multicolumn{2}{c|}{Ours}           & \begin{tabular}[c]{@{}c@{}}{94.64±0.59}\end{tabular} & \begin{tabular}[c]{@{}c@{}}{90.32±0.67}\end{tabular} & \begin{tabular}[c]{@{}c@{}}{91.60±0.66}\end{tabular}  & \begin{tabular}[c]{@{}c@{}}{93.90±0.78}\end{tabular} & \begin{tabular}[c]{@{}c@{}}{66.44±1.02}\end{tabular}   & \begin{tabular}[c]{@{}c@{}}{82.52±1.10}\end{tabular}         & \begin{tabular}[c]{@{}c@{}}{65.34±1.67}\end{tabular}   & \begin{tabular}[c]{@{}c@{}}{85.14±0.71}\end{tabular} \\ \bottomrule \bottomrule
\end{tabular} 
\end{table*}

\begin{table*}
\renewcommand{\arraystretch}{1.5}
\centering
\footnotesize
\caption{Comparison of Performances (\%) in MSTAR}
\label{comparisonMSTAR1}
\begin{tabular}{c|c|cccccccccc}
\toprule \toprule 
\multicolumn{2}{c|}{\multirow{2}{*}{Algorithms}} & \multicolumn{10}{c}{Image Number for Each Class}                                                         \\ \cline{3-12} 
\multicolumn{2}{c|}{}                            & \multicolumn{1}{c}{10}  &  \multicolumn{1}{c}{20} &  \multicolumn{1}{c}{30} & \multicolumn{1}{c}{40} & \multicolumn{1}{c}{55} &\multicolumn{1}{c}{80} &\multicolumn{1}{c}{110} &\multicolumn{1}{c}{165} &\multicolumn{1}{c}{220} & \multicolumn{1}{c}{All data} \\ \midrule
\multirow{6}{*}{\begin{tabular}[c]{@{}c@{}}Data\\augmentation-based \\for limited-data \\SAR ATR\end{tabular}} &GAN-CNN1 \cite{comparison1}       &-       & 81.80 &-            & 88.35      & -      & 93.88    &- &- &-         & 97.03                 \\
& GAN-CNN2 \cite{comparison1}    &-    & 84.39   &-          & 90.13    &   -      & 94.91   &- & -& -         & 97.53                 \\ 
&MGAN-CNN \cite{comparison1}      &-      & 85.23     &-        & 90.82    & -        & 94.91   & -&- & -         & 97.81                 \\ 
&Triple-GAN\cite{addc3}   &-     &-&-    &-           & 95.70      &-     & 95.97           & 96.13                   & 96.46      &-        \\
&Improved-GAN     \cite{addc4}   &-  &-&-     &-        & 87.52        &-   & 95.02           & 97.26                     & 98.07    &-       \\
&Semi-supervised GAN\cite{addc5}   &-  &-&-    &-        & 95.72      &-     & 97.22           & 97.97            & 98.14         &-    \\
&Supervised \cite{reduce1}  &-   &92.62  &- &97.11            &-      &98.65     &-           &-            &-         &-    \\ \midrule 
\multirow{11}{*}{\begin{tabular}[c]{@{}c@{}}New model/module \\for limited-data \\SAR ATR\end{tabular}} &Deep CNN \cite{comparison2}         &-        & 77.86     &-        & 86.98     &     -   & 93.04    &- &- &-         & 95.54                 \\
&Improved DNN \cite{addnew8}     &-            & 79.39  &-           & 87.73    &  -       & 93.76    &- &- &  -       & 96.50                 \\
&Dens-CapsNet\cite{ac2}      &80.26       & 92.95    &-         & 96.50      &    -       & -      & -&- &    -       & 99.75                     \\ 
& HDLM\cite{ac9}      &88.16      & 95.17     &-       & 97.85      &    -       & 98.80      & -&- &    -       & -                     \\ 
& ASC-MACN\cite{ac10}      &62.85      & 79.46      &-      & -      &    -       & -      & -&- &    -       & 99.42                     \\ 
& ASC-MBCRN\cite{ac10}      &87.96      & 96.04     &-       & -      &    -       & -      & -&- &    -       & 99.96                     \\
& DBRSA\cite{lv2023simulation}      &77.36      & -     &92.42       & -      &    97.58(50)       & -      & -&- &    -       & -                     \\
& PD Network\cite{zhang2024optimal}      &83.73      & 93.00     &94.63       & -      &    -       & -      & -&- &    -       & -                     \\
\midrule 
\multicolumn{2}{c|}{Ours   }                      & 92.25     & 96.37  &96.78  & 97.40 & -  & 99.67  &- &- & 99.96 & -                     \\ \bottomrule \bottomrule
\end{tabular}
\end{table*}

\subsubsection{Comparison under OpenSARShip}\label{sec: Comparison under OpenSARShip}

This subsection compares the performance of the proposed method with several state-of-the-art approaches using a constant number of training samples.

The quantitative comparison with other modern deep learning-based approaches is shown in Table~\ref{3&6COMPARISONOpenSARShip}, evaluating four key metrics: recall, precision, F1-score, and accuracy. For the three-class and six-class recognition task, our method achieves superior performance. These results outperform the best results from other methods.

These comparisons clearly demonstrate that our proposed method achieves state-of-the-art recognition performance across different training sample sizes. It consistently outperforms all other techniques in both types of recognition tasks, whether involving varying class numbers or different amounts of training data.

\subsubsection{Comparison under MSTAR}\label{sec: Comparison under MSTAR}

To evaluate the impact of limited SAR training samples, we compare our method with state-of-the-art approaches using varying amounts of training data, from the full dataset to as few as 40 samples, as shown in Table \ref{comparisonMSTAR1}.

We randomly sampled $K$ shots per class from the MSTAR dataset and categorized the methods into three groups: data augmentation-based methods (GAN-CNN1, GAN-CNN2, MGAN-CNN, Triple GAN, Improved GAN, Semi-supervised GAN, Supervised ATR), and deep learning-based methods introducing novel models or modules (deep CNN, Improved DNN, Simple CNN, Metric Learning).

As shown in Table \ref{comparisonMSTAR1}, our method achieves state-of-the-art recognition performance, comparable to other SAR target recognition algorithms. Additionally, it not only delivers accurate results but also provides interpretable reasoning for the predictions.

These results confirm the effectiveness and rationality of our approach, demonstrating its strong generalization ability across various network architectures and experimental conditions, which highlights its potential for practical applications.

\section{Conclusion}
\label{conclusions}

This paper addresses the challenges posed by the black-box nature of deep learning in SAR ATR by proposing a physics-based two-stage feature decomposition method. 
This method effectively transforms uninterpretable deep features into ASCCs with clear physical meaning, enhancing both the interpretability and accuracy of SAR ATR methods. 
By introducing an FDD module and designing a MLO-NMTF with targeted orthogonal constraints, the proposed approach decouples ASCCs from deep features during pre-decomposition and then decomposes ASCCs with different physical meanings from deep features layer by layer.
The designed experiments first validate the intra-class compactness and inter-class separation of ASCCs, ensuring that they serve as reasoning bases for interpretable reasoning while maintaining high discriminative power for accurate recognition. Additionally, the experiments confirm the method's low error distribution in decomposing deep features into ASCCs, demonstrating the effectiveness of the decomposition. Finally, the approach proves capable of achieving accurate recognition across various common deep-learning architectures and under conditions of limited SAR training data and limited ASCC numbers.
This work aims to facilitate the practical deployment of deep learning-based SAR ATR in decision-critical applications, where interpretability and reliability are important.

\bibliographystyle{IEEEtran}
\bibliography{bib/ASC}

\end{document}